\documentclass[12pt]{article}
\usepackage{graphicx}
\usepackage{amsfonts}
\usepackage{amssymb,amsmath}
\usepackage{latexsym}
\usepackage{color}
\usepackage{cite}
\usepackage{bm}
\input{colordvi.tex}

\renewcommand{\thefootnote}{\#\arabic{footnote}}

\begin{document}

\renewcommand{\thepage}{\arabic{page}}
\setcounter{page}{1}
\renewcommand{\thefootnote}{\#\arabic{footnote}}

\begin{titlepage}
\begin{center}

\begin{flushright}
ICRR-Report-686-2014-12\\
\end{flushright}

\vskip .5in
{\Large \bf 
Blue-tilted Tensor Spectrum  \\ and Thermal History of the Universe
}

\vskip .45in

{\large
Sachiko~Kuroyanagi$^1$,
Tomo~Takahashi$^2$ and 
Shuichiro~Yokoyama$^3$
}

\vskip .45in

{\em
$^1$
Department of Physics, Faculty of Science, Tokyo University of Science, 1-3, Kagurazaka,
Shinjuku-ku, Tokyo 162-8601, Japan  \vspace{2mm} \\
$^2$
Department of Physics, Saga University, Saga 840-8502, Japan \vspace{2mm} \\
$^3$
Institute for Cosmic Ray Research, University of Tokyo, Kashiwa 277-8582, Japan \vspace{2mm} \\
}

\end{center}

\vskip .4in

\begin{abstract}
  We investigate constraints on the spectral index of primordial
  gravitational waves (GWs), paying particular attention to a
  blue-tilted spectrum.  Such constraints can be used to test a
  certain class of models of the early Universe.  We investigate
  observational bounds from LIGO+Virgo, pulsar timing and big bang
  nucleosynthesis, taking into account the suppression of the
  amplitude at high frequencies due to reheating after inflation and
  also late-time entropy production.  Constraints on the spectral
  index are presented by changing values of parameters such as
  reheating temperatures and the amount of entropy produced at late
  time.  We also consider constraints under the general modeling
  approach which can approximately describe various scenarios of the
  early Universe.  We show that the constraints on the blue spectral
  tilt strongly depend on the underlying assumption and, in some
  cases, a highly blue-tilted spectrum can still be allowed.
\end{abstract}
\end{titlepage}

\setcounter{footnote}{0}

\section{Introduction}

Primordial gravitational waves (GWs) are one of the most important
probe of the very early Universe and a lot of efforts have been made
for this subject both in theoretical and observational aspects.  In
particular, the recent result of B-mode polarization in cosmic
microwave background (CMB) from BICEP2 \cite{Ade:2014xna,Ade:2014gua}
that claimed the detection of the primordial GWs has been stimulating
a lot of research.  Although there have been some debate regarding the
issue of the foreground subtraction in BICEP2
\cite{Liu:2014mpa,Mortonson:2014bja,Flauger:2014qra}, taking it at
face value, the BICEP2 result seems to be inconsistent with the
temperature data from Planck.  From the analysis of Planck+WMAP 9-year
polarization in the framework of the $\Lambda$CDM+tensor mode, the
upper bound on the tensor-to-scalar ratio $r$ has been given as $r <
0.11$ (95~\% C.L) \cite{Ade:2013uln}, while the recent BICEP2 gives
the bound $r = 0.2^{+0.07}_{-0.05}$ (68 \% C.L.) \cite{Ade:2014xna}.
This tension motivates various works considering an extension of the
standard cosmological model, one of which is a blue-tilted tensor
spectrum
\cite{Gong:2014qga,Gerbino:2014eqa,Wang:2014kqa,Ashoorioon:2014nta,Wu:2014qxa}.
Although the standard inflation model predicts a red-tilted spectrum
since the tensor spectral index $n_T$ has the so-called consistency
relation $n_T = - r /8$, a blue-tilted spectrum can be realized in
some models, e.g, string gas cosmology \cite{Brandenberger:2006xi},
super-inflation models \cite{Baldi:2005gk}, G-inflation
\cite{Kobayashi:2010cm}, non-commutative inflation
\cite{Calcagni:2004as,Calcagni:2013lya}, particle production during
inflation \cite{Cook:2011hg,Mukohyama:2014gba}, and so on.  Therefore,
observational constraints on the blue-tilted tensor spectrum would be
worth investigating also from the perspective of models of the early
Universe.

Aside from the CMB, primordial GWs could be also detected as a
stochastic GW background and there are several observational
constraints on the energy density of the stochastic GWs at different
GW frequencies: pulsar timing
\cite{Jenet:2006sv,vanHaasteren:2011ni,Demorest:2012bv,Zhao:2013bba},
big bang nucleosynthesis (BBN) \cite{Allen:1996vm,Maggiore:1999vm},
interferometric GW detectors such as LIGO and Virgo
\cite{Aasi:2014zwg} and so on~\footnote{
  There are more ways to obtain upper bounds on the amount of the
  stochastic GWs, such as dark radiation constraints from the CMB
  \cite{Smith:2006nka,Sendra:2012wh}, CMB $\mu$ distortion
  \cite{Ota:2014hha,Chluba:2014qia}, helioseismology
  \cite{Siegel:2014yta}, precision Doppler tracking from the Cassini
  spacecraft \cite{Armstrong:2003ay}, orbital monitoring of binary
  systems \cite{Hui:2012yp}, torsion-bar antennas
  \cite{Shoda:2013oya}, seismic spectrum from the Earth
  \cite{Coughlin:2014sca}, synchronous recycling interferometers
  \cite{Akutsu:2008qv}, cross-correlation measurement between the
  Explorer and Nautilus cryogenic resonant bar detectors
  \cite{Astone:1999a a}, and Global Positioning System (GPS) satellite
  \cite{Aoyama:2014fea}.
}.  Although these limits are far above the prediction of the standard
inflationary models, a strongly blue-tilted tensor spectrum can easily
be excluded by these constraints, which has been investigated in
\cite{Stewart:2007fu,Camerini:2008mj}.

However, it should be noted that the thermal history of the Universe
affects the spectrum of the primordial GWs
\cite{Seto:2003kc,Boyle:2007zx,Nakayama:2008ip,Nakayama:2008wy,Kuroyanagi:2009br,Kuroyanagi:2011fy,Kuroyanagi:2013ns,Kuroyanagi:2008ye,Nakayama:2009ce}.
For instance, high-frequency primordial GWs which entered the horizon
during a matter-like component dominated epoch, such as that exists
during reheating or late-time entropy production, have different
frequency dependence compared to those which entered the radiation
dominated epoch.  This results in a suppression of the spectrum at
high frequencies, whose range depends on when and how long it took
place.  It should also be noted that the above-mentioned generation
mechanisms of GWs do not necessarily predict a blue-tilted spectrum
over all the frequencies.  The spectral index of the primordial GW
spectrum can change at some frequency.  Therefore, if we also take
into account the spectral shapes caused by those effects, the
constraint on $n_T$ changes depending on the underlying scenario of
the early Universe, which we are going to investigate in this paper.

For this purpose, we use two different approaches to describe the GW
spectrum.  First, we consider constraints on $n_T$ taking into account
the suppression of the spectrum at high frequencies due to reheating
and late-time entropy production, assuming that the primordial
spectrum has uniform spectral index over all frequencies.  For this
case, we use a fitting formula which can reproduce the effect of the
thermal history on the spectrum with a very good accuracy.  Second, we
consider the constraints on $n_T$ without assuming an explicit model
of the early Universe.  Since the shape of the GW spectrum strongly
depends on the model assumed, we use a more general form of the
spectrum such that the spectral index changes from $n_T$ to a
different value at a given frequency.  This modeling also includes the
case where the Universe is dominated by a component whose equation of
state differs from that of radiation/matter component.  Although we
assume that the transition is discontinuous, this method can
approximately describe GW spectra in several scenarios of the early
Universe.

This paper is organized as follows.  In the next section, we begin
with a brief review of the formalism to calculate the GW spectrum,
taking into account the effect of thermal history by using fitting
functions.  In Section~\ref{sec:results}, first we summarize current
observational bounds on the amplitude of the GW to be used in
providing an upper bound on $n_T$.  Then we show how the constraints
on $n_T$ change depending on the existence of reheating and late-time
entropy production.  Subsequently, we provide constraints on $n_T$
under the general modeling.  The final section is devoted to the
conclusion of this paper.

\section{GW spectrum and thermal history of the Universe}  \label{sec:spectrum}
Here, we briefly summarize the formalism to calculate the spectrum of
a stochastic GW background of primordial origin.  GWs are described as
a tensor part of the metric perturbation in the flat
Friedmann-Robertson-Walker (FRW) background, which is given as
\begin{equation}
ds^2 = -dt^2 + a^2(t)\left( \delta_{ij} + h_{ij} (t,{\bm x})\right) dx^i dx^j  = a^2(\tau)\left[ - d\tau^2 + \left( \delta_{ij} + h_{ij} (\tau,{\bm x})\right) dx^i dx^j \right],
\end{equation}
where $a(\tau)$ is the scale factor and $\tau$ is the conformal time
which is related to the cosmic time $t$ with $d\tau = dt / a$.  The
tensor metric perturbation $h_{ij}$ satisfies the transverse-traceless
condition $\partial^i h_{ij} = h^i_{~i} = 0$.  The energy density of
GWs $\rho_{\rm GW}$ is given by
\begin{equation}
\rho_{\rm GW} = \frac{1}{64 \pi G a^2} \left\langle  \left( \partial_\tau h_{ij}  \right)^2+  \left( \nabla h_{ij}  \right)^2 \right\rangle,
\end{equation}
where the bracket indicates the spacial average.  By Fourier
transforming $h_{ij} (\tau,{\bm x})$ as
\begin{equation}
h_{ij} (\tau, {\bm x}) = \sum_{\lambda = +, \times} \int \frac{dk^3}{(2\pi)^{3/2}} \epsilon_{ij}^{\lambda} h_{\bm k} (\tau) e^{i {\bm k}\cdot {\bm x}},
\end{equation}
with $\epsilon_{ij}^\lambda$ being the polarization tensor, which
satisfies the symmetric, transverse-traceless condition and normalized
by the relation $\sum_{i,j} \epsilon_{ij}^{\lambda} \left(
  \epsilon_{ij}^{\lambda'} \right)^\ast = 2 \delta^{\lambda\lambda'}$,
the GW energy density $\rho_{\rm GW}$ can be rewritten as
\begin{equation}
\label{eq:rho_GW}
\rho_{\rm GW} 
= \frac{1}{32 \pi G}  \int d \ln k \left( \frac{k}{a} \right)^2 \frac{k^3}{\pi^2} \sum_\lambda \left| h_{\bm k}^\lambda \right|^2.
\end{equation}

Conventionally, the amplitude of GWs is characterized in the form of
the energy density parameter of GWs per logarithmic interval of the
wave number $k$~\footnote{
  In the later part of the paper, we also use frequency $f=k/2\pi$
  instead of $k$.
} 
normalized by the critical density $\rho_{\rm crit} (t) = 3H^2 /(8\pi G)$, 
\begin{equation}
\Omega_{\rm GW} (k) \equiv \frac{1}{\rho_{\rm crit}} \frac{ d \rho_{\rm GW}}{d \ln k}.
\end{equation}
Using
Eq.~\eqref{eq:rho_GW}, we can express $\Omega_{\rm GW}$ as 
\begin{equation}
\label{eq:Omega_GW}
\Omega_{\rm GW} (k)  = \frac{1}{12} \left( \frac{k}{aH} \right)^2 {\cal P}_T (k).
\end{equation}
Here we have introduced the power spectrum $\mathcal{P}_T(k)$ as 
\begin{equation}
\mathcal{P}_T (k) = \frac{k^3}{\pi^2} \sum_\lambda \left| h_{\bm k}^\lambda \right|^2 =   T_T^2 (k) \mathcal{P}^{\rm prim}_T (k),
\end{equation}
where $T_T (k)$ is the transfer function and $\mathcal{P}^{\rm prim}_T
(k)$ is the primordial tensor power spectrum.  $\mathcal{P}^{\rm
  prim}_T(k)$ is commonly parametrized as
\begin{equation}
\mathcal{P}^{\rm prim}_T (k) = A_T (k_{\rm ref} ) \left( \frac{k}{k_{\rm ref}} \right)^{n_T},
\end{equation}
where $A_T (k_{\rm ref})$ and $n_T$ are the amplitude and the spectral
index at the reference scale $k_{\rm ref}$.  The amplitude, $A_T
(k_{\rm ref})$, is usually characterized by the so-called
tensor-to-scalar ratio $r$ defined by
\begin{equation}
r (k_{\rm ref} )  \equiv  \frac{\mathcal{P}^{\rm prim}_T (k_{\rm ref} )}{\mathcal{P}_\zeta (k_{\rm ref})},
\end{equation}
and we have
\begin{eqnarray}
A_T(k_{\rm ref}) = r {\mathcal P}_\zeta (k_{\rm ref}),
\end{eqnarray}
where $\mathcal{P}_\zeta (k_{\rm ref})$ is the power spectrum for the
scalar perturbation and it is precisely measured as $\mathcal{P}_\zeta
= 2.2 \times 10^{-9}$ at $k_{\rm ref}=0.01{\rm Mpc}^{-1}$
\cite{Ade:2013uln}.

The transfer function $T_T (k)$ describes the evolution of GWs after
horizon crossing, which depends on the thermal history of the
Universe.  This can be obtained by numerically solving the evolution
equation of the GWs:
\begin{equation}
\label{eq:EOM_h}
{h}_{ij}'' + 2 a H {h}_{ij}' -  \nabla^2 h_{ij} = 0,
\end{equation}
where a prime denotes a derivative with respect to the conformal time.
In the following subsections, we provide fitting formulas which enable
us to easily express the effects of thermal history with a few
parameters.

\subsection{Standard reheating scenario}
Let us first consider the standard reheating scenario in which soon
after inflation the Universe has matter-dominated (MD) like phase
before the completion of reheating. Such a phase is proceeded by the
oscillation of inflaton field at the bottom of its quadratic
potential.  After the completion of reheating, the Universe becomes
radiation-dominated (RD).  In such a case, the transfer function is
described as
\cite{Turner:1993vb,Nakayama:2008wy,Nakayama:2009ce,Kuroyanagi:2011fy,Chongchitnan:2006pe}:
\begin{equation}
T_T^2 (k) =  
\Omega_m^2 \left( \frac{g_\ast ( T_{\rm in})}{g_{\ast 0}} \right)
 \left( \frac{g_{\ast s0}}{g_{\ast s} (T_{\rm in})} \right)^{4/3}
 \left( \frac{ 3 j_1 (k\tau_0)}{k \tau_0} \right)^2 T_1^2 (x_{\rm eq}) T_2^2 (x_R),
\end{equation} 
where $j_\ell (k \tau_0)$ is the $\ell$th spherical Bessel function,
given by $ j_1 (k \tau_0) = 1 / (\sqrt{2} k \tau_0)$ in the limit of $
k \tau_0 \rightarrow 0$.  The subscript ``0" indicates that the
quantity is evaluated at the present time.  The values of the
relativistic degrees of freedom $g_\ast ( T_{\rm in})$ and its
counterpart for entropy $g_{\ast s} (T_{\rm in})$ change depending on
$T_{\rm in}(k)$, which is the temperature of the Universe when the
mode $k$ enters the horizon, and affects the spectral amplitude of the
mode \cite{Watanabe:2006qe}\footnote{
  To incorporate this effect in the GW spectrum, we introduce the
  following fitting function: 
\begin{equation*}
g_\ast (T_{\rm in}(k))=g_{\ast 0} \left\{ 
\frac{A+\tanh \left[-2.5\log_{10}\left( \frac{k/2\pi}{2.5\times 10^{-12} ~{\rm Hz}}\right) \right]}{A+1}
\right\}
\left\{
\frac{B+\tanh \left[-2.0\log_{10}\left( \frac{k/2\pi}{6.0\times 10^{-9} ~{\rm Hz}} \right) \right]}{B+1}
\right\} 
\end{equation*} 
where $A=(-1-10.75 /g_{\ast 0})/(-1+10.75/g_{\ast 0})$ and
$B=(-1-g_{\rm max}/10.75) /(-1+g_{\rm max}/10.75)$.  For $g_{\rm
  max}$, we assume the sum of the standard-model particles, $g_{\rm
  max}=106.75$.  The same formula can be used for the counterpart for
entropy $g_{\ast s}(T_{\rm in})$ by replacing $g_{\ast 0}=3.36$ with
$g_{\ast s0}=3.91$.
}.  The index $i$ in $x_i\equiv k / k_i$ corresponds to the transition
epoch of the Hubble expansion rate.  The wavenumbers corresponding to
the matter-radiation equality and the completion of reheating are
respectively given by
\begin{eqnarray}
\label{eq:k_eq}
k_{\rm eq} 
 &=& 7.1 \times 10^{-2} \Omega_m h^2~{\rm Mpc}^{-1}, \\
\label{eq:k_R}
k_R &=&  1.7 \times 10^{14}  \left( \frac{g_{\ast s} (T_R) }{106.75} \right)^{1/6} \left( \frac{T_R}{10^7~{\rm GeV}} \right) ~{\rm Mpc}^{-1}.
\end{eqnarray}
For $T_1^2(x)$ and $T_2^2(x)$, we use the following fitting functions,
\begin{eqnarray}
T_1^2(x)  & = &  1 + 1.57 x + 3.42 x^2, \\
T_2^2(x)  & = &  (1 - 0.22 x^{1.5} + 0.65 x^2)^{-1}.
\end{eqnarray}
Note that the fitting formula for $T_2^2(x)$ is improved from
\cite{Nakayama:2008wy} by changing the power of the power index of the
second term.  Using least-square algorithm, the power index and
coefficients are readjusted to fit the spectrum obtained by
numerically solving Eq.~(\ref{eq:EOM_h}).  (For the details of the
numerical calculation, see \cite{Kuroyanagi:2008ye}.)~\footnote{
  Note that, since $x_R$ includes $T_R$, the coefficients will be
  modified depending on the definition of $T_R$ in terms of the decay
  rate of the inflaton $\Gamma$.  Numerical results are obtained for a
  given value of $\Gamma$, not $T_R$, and we relate these quantities
  by using the formula
  $T_R=(10/\pi^2)^{1/4}g_*(T_R)^{-1/4}\sqrt{\Gamma M_{\rm pl}}$ with
  $M_{\rm pl} = (8 \pi G)^{-1/2}$ being the reduced Planck mass.
}

\subsection{Late-time entropy production scenario}
We also consider the case with late-time entropy production, in which
the oscillation energy of another scalar field $\sigma$ dominates the
Universe some time after reheating.  In some cases, during the RD era
after the decay of inflaton, the oscillating $\sigma$ field can
dominate over the radiation energy density.  Then, the Universe enters
a MD-like phase due to the $\sigma$ oscillation, ~\footnote{
  When the energy density of the $\sigma$-field starts to dominate the
  Universe, the $\sigma$-field is not necessarily oscillating.  If the
  field is frozen by the Hubble friction, the Universe experiences a
  second inflationary phase and the amplitude of GWs at the scales
  which enter the horizon during this phase, is suppressed more
  quickly than in the case of MD-like phase\cite{Mendes:1998gr}.
  Here, we focus on only MD-like phase during $\sigma$-dominated
  Universe, but it would be easy to extend our analysis to the case
  with such a second inflationary phase.
}and after the $\sigma$-field decays into radiation, the Universe is
dominated by the radiation energy again.  For this case, the fitting
formula for the transfer function is given by \cite{Nakayama:2009ce}
\begin{equation}
\label{eq:T_T_entropy}
T_T^2 (k) =  
\Omega_m^2 \left( \frac{g_\ast ( T_{\rm in})}{g_{\ast 0}} \right)
 \left( \frac{g_{\ast s0}}{g_{\ast s} (T_{\rm in})} \right)^{4/3}
 \left( \frac{ 3 j_l (k\tau_0)}{k \tau_0} \right)^2 
 T_1^2 (x_{\rm eq}) T_2^2 (x_\sigma) T_3^2 (x_{\sigma R}) T_2^2 (x_{RF}),
\end{equation}
where $k_\sigma$ corresponds to the time when $\sigma$ decays into
radiation after $\sigma$-dominated MD-like phase, and given in the
same form as Eq.~\eqref{eq:k_R},
\begin{eqnarray}
k_\sigma &=&  1.7 \times 10^{14}  \left( \frac{g_{\ast s} (T_\sigma) }{106.75} \right)^{1/6} \left( \frac{T_\sigma}{10^7~{\rm GeV}} \right) ~{\rm Mpc}^{-1},
\end{eqnarray}
with $T_\sigma$ being the temperature of the Universe at the $\sigma$
decay.  To write down $k_{\sigma R}$ and $k_{RF}$, first we define the
quantity which represents the amount of entropy production by the
decay of $\sigma$ as
\begin{equation}
\label{eq:def_F}
F \equiv \frac{ s (T_\sigma) a^3 (T_\sigma) }{s (T_R) a^3 (T_R) },
\end{equation}
where $s(T)$ is the entropy density at temperature $T$. 
With this quantity, the other characteristic frequencies can be
expressed as $k_{\sigma R} = k_\sigma F^{2/3}$ and $k_{RF} = k_R
F^{-1/3}$.  The third transfer function $T_3(x)$ describes the
transition from the first RD phase to $\sigma$-dominated MD-like phase
and we use
\begin{eqnarray}
T_3^2(x)  & = &  1 + 0.59 x + 0.65 x^2.
\end{eqnarray}
Note that $T_1(x)$ is used instead of $T_3(x)$ in
\cite{Nakayama:2009ce}.  In this paper, we introduce a new function
$T_3(x)$ in order to improve the fitting.

\begin{figure}[htbp]
  \begin{center}
     \includegraphics[width=0.49\textwidth]{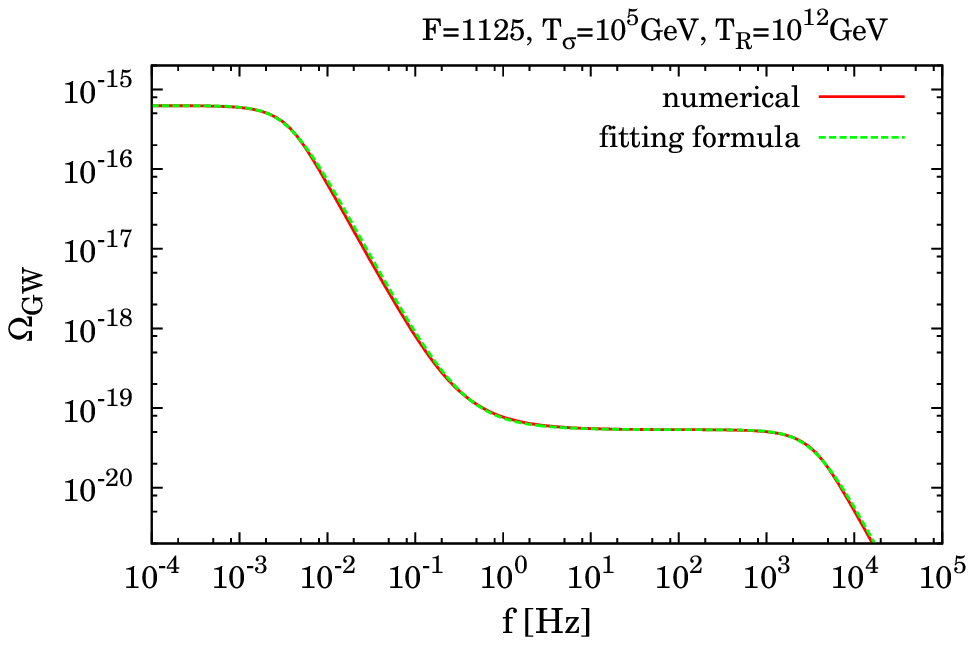}
     \includegraphics[width=0.49\textwidth]{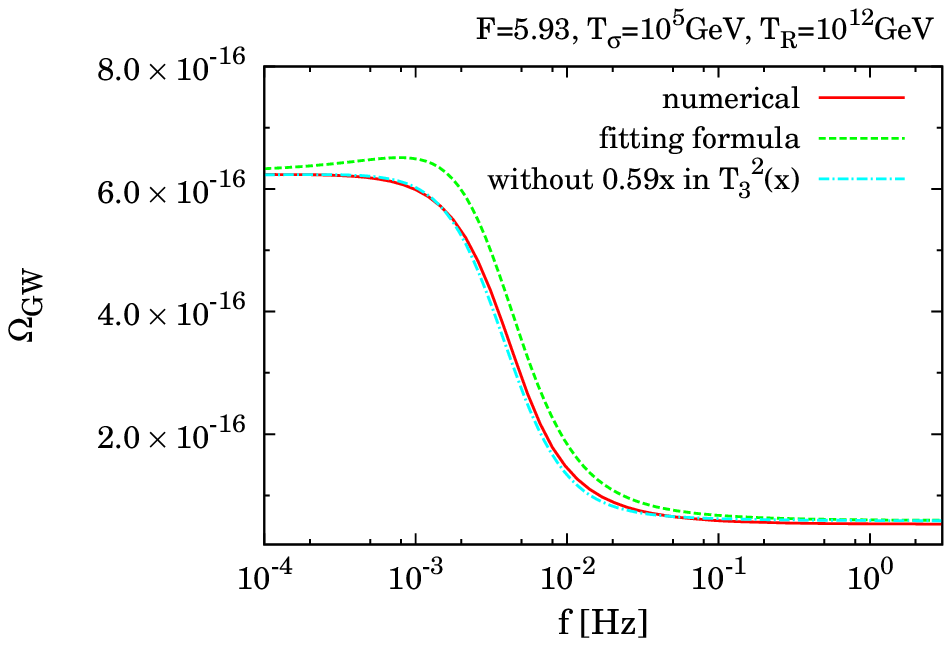}
   \end{center}
   \caption{
     \label{fig:fitting} Comparison of spectra derived from the
     fitting formula~\eqref{eq:T_T_entropy} and numerical calculation
     for $F=1125$ (left) and $F=5.93$ (right).  In the right panel,
     the spectra are plotted in linear scale and we also show the
     spectrum using the fitting function without the term $0.59x$ of
     $T_3(x)$ in Eq.~\eqref{eq:T_T_entropy}, which causes an
     artificial bump for small $F$.  In both figures, we
     assume  $r=0.2$ and $n_T=0$.}
\end{figure}

In Fig.~\ref{fig:fitting}, we show the spectra from the fitting
formula~\eqref{eq:T_T_entropy}, and the numerical calculation.
As seen from the figure, our fitting formula well approximates the one
from the numerical calculation.  We note here that our new transfer
functions $T_2(x)$ and $T_3(x)$ are also aimed to reduce a small bump
around the bending point, which arises in the fitting formulas of
\cite{Nakayama:2009ce} for small $F$.  However, even with the improved
formulas, as seen in the right panel of Fig.~\ref{fig:fitting}, a bump
is unavoidable for $F\lesssim 10$ because the frequency transitions
described by $T_2(x)$ and $T_3(x)$ overlap with each other in a narrow
frequency range.  The bump can be removed by omitting the second term
of $T_3(x)$, namely $0.59 x$ as shown in the right panel of
Fig.~\ref{fig:fitting}.  In the following analysis, we omit this term
for $F \lesssim 10$.

\section{ Constraints on the tensor spectral index} \label{sec:results}
\subsection{Current observational bounds on GWs}
First, we summarize observational bounds on the energy density of the
stochastic GWs, which are used to obtain constraints on $n_T$.  Among
various upper bounds reported in the literature, we adopt stringent
ones from interferometric GW detectors, pulsar timing array, and BBN.

\subsubsection{Interferometric GW detectors such as LIGO+Virgo} 
From interferometric GW detectors such as LIGO, non-detection of GWs
gives an upper bound on the stochastic GWs.  In this paper, we adopt
the 95~\% C.L. upper bound from the LIGO-Virgo collaboration
\cite{Aasi:2014zwg}\footnote{
  In \cite{Aasi:2014zwg}, the bounds on the GW amplitude for the
  frequency regions $f = 600$--$1000$~Hz and $1000$--$1726$~Hz are
  also given.  Although these bounds may be able to constrain very
  blue-tilted spectrum, we checked that these bounds are irrelevant in
  the parameter regions which we explore in our analysis.  Thus we
  only consider the bound for $f = 41.5$--$169.25$~Hz in the following
  analysis.
}:
\begin{equation}
\label{eq:LIGO}
\Omega_{\rm GW}h^2 < 2.6 \times 10^{-6} 
\qquad \left[ f= 41.5 -169.25~{\rm Hz} \right],
\end{equation}
where $h$ is the dimensionless Hubble parameter which parametrizes the
present Hubble constant as $H_0=100 h {\rm km/s/Mpc}$.  The analysis
is performed using the approximate form of the GW spectrum,
$\Omega_{\rm GW}(f)=\Omega_{\rm GW,\alpha}(f/100{\rm Hz})^\alpha$,
where $\alpha$ is the local power index around the sensitive frequency
band $\sim 100$Hz, and Eq. (\ref{eq:LIGO}) is obtained assuming
$\alpha=0$.  The limit changes for different values of $\alpha$, and
we describe its $\alpha$ dependence as
\begin{equation}
\label{eq:LIGO2}
\Omega_{\rm GW,\alpha}h^2< 2.6 \times 10^{-6}\sqrt{\frac{5-2\alpha}{5}}\left(\frac{f_{\rm min}}{100{\rm Hz}}\right)^{-\alpha}.
\end{equation}
This formula is obtained by adopting the approximations described
below.  The signal-to-noise ratio for the analysis of stochastic GWs
is given by \cite{Abadie:2011fx}
\begin{equation}
SNR^2=\left(\frac{3H_0^2}{10\pi^2}\right)^2 2T_{\rm obs}\int^{f_{\rm max}}_{f_{\rm min}} df \frac{\gamma_{ij}^2(f)\Omega_{\rm GW}^2(f)}{f^6P_i(f)P_j(f)},
\end{equation} 
where $T_{\rm obs}$ is the observation time.  Assuming that the
overlap reduction function $\gamma_{ij}^2(f)$ and the noise power
spectral density $P_{i,j}(f)$ have no frequency dependence, which is a
reasonable assumption between $f_{\rm max}=41.5$~Hz and $f_{\rm
  min}=169.25$~Hz, we obtain $SNR^2\sim A^2 \Omega_{\rm
  GW,\alpha}^2(f_{\rm min}/100{\rm Hz})^{2\alpha}/[(5-2\alpha)f_{\rm
  min}^5]$, where $A^2\equiv (3H_0^2/10\pi^2)^22T_{\rm obs}
\gamma_{ij}^2/(P_iP_j)$ and we have used $f_{\rm min}^{2\alpha-5}\gg
f_{\rm max}^{2\alpha-5}$ for $2\alpha-5<0$.  This approximation holds
in our investigation where the value of $\alpha$ ranges from $-2$ to
$1.6$.  Assuming a Gaussian likelihood, the bound is $\Omega_{\rm
  bound,\alpha}\propto\Delta\Omega_{\rm GW,\alpha}\propto 1/SNR$
\cite{Seto:2005qy}.  Using that the bound for $\alpha=0$ is
$\Omega_{\rm bound,0}\propto \sqrt{5f_{\rm min}^5}/A$, we obtain
\begin{equation}
\Omega_{\rm bound,\alpha}={\sqrt{(5-2\alpha)f_{\rm min}^5} \over A}\left({f_{\rm
  min} \over 100{\rm Hz}} \right)^{-\alpha}=\Omega_{\rm
  bound,0} \sqrt{{5-2\alpha \over 5}}\left({f_{\rm min} \over 100{\rm Hz}}\right)^{-\alpha},
\end{equation}
which gives Eq. (\ref{eq:LIGO2}).

\subsubsection{Pulsar timing array}
We also adopt the upper bound from pulsar timing array.  Using
millisecond pulsars as very precise clocks, GWs can be searched
through the effect on the pulse arrival timings, which currently
provides the stringent constraint on the amplitude of GWs at $f\sim
10^{-8}$ Hz.  Among several bounds reported so far, we use the recent
one obtained from the North American Nanohertz Observatory for
Gravitational waves (NANOGrav) project \cite{Demorest:2012bv}, which
gives the upper bound:
\begin{equation}
\Omega_{\rm GW} h^2 < 1.1 \times 10^{-8} \qquad \left[ f= 1/  (5.54~{\rm years}) = 5.72 \times 10^{-9}{\rm Hz} \right],
\label{eq:NANOGrav}
\end{equation}
In \cite{Demorest:2012bv}, GW spectrum is modeled by $h_c(f)=A_{\rm
  1year}(f/f_{\rm 1year})^\beta$, which corresponds to the energy
density as $\Omega_{\rm GW}(f)=(2\pi^2/3H_0^2)f^2h_c^2(f)$, and the
spectral index dependence of the upper bound is found to be well
approximated by $A_{\rm 1year}=2.26\times 10^{-14}(5.54{\rm
  year}/1{\rm year})^\beta$.  This dependence can be canceled by
choosing the reference frequency to be $f=1/(5.54{\rm years})=5.72
\times 10^{-9}{\rm Hz}$.  Then the bound becomes
\begin{equation}
  h_c(f_{\rm 5.54year})< 2.26\times 10^{-14} \left(\frac{5.54{\rm year}}{1{\rm year}}\right)^\beta \left(\frac{f_{\rm 5.54year}}{f_{\rm 1year}}\right)^\beta=2.26\times 10^{-14},
\end{equation}
from which we obtain the bound Eq.~\eqref{eq:NANOGrav}.

\subsubsection{Big Bang Nucleosynthesis (BBN)}
We also consider constraints from BBN.  Primordial GWs contribute to
the energy density of the Universe as an extra radiation component.
Such an extra radiation component changes the expansion rate of the
Universe during BBN and affects the abundance of the light elements.
The total energy density of GWs, given by integrating the density
parameter $\Omega_{\rm GW}(f)$, is therefore constrained not to spoil
BBN \cite{Allen:1996vm,Maggiore:1999vm}:
\begin{equation}
\label{eq:BBN}
\int_{f_1}^{f_2} d (\ln f) \Omega_{\rm GW} (f) h^2 \le 5.6 \times 10^{-6} (N_{\rm eff}^{\rm (upper)} -3),
\end{equation}
where $N_{\rm eff}^{\rm (upper)}$ is the upper bound on the effective
number of extra radiation at the time of BBN.  The lower limit of the
integral in the left hand side is given by the frequency which
corresponds to the comoving horizon at the time of BBN and we take
$f_1 = 10^{-10}~{\rm Hz}$.  For the upper cutoff, we take $f_2 =
10^7~{\rm Hz}$, which corresponds to the temperature of the Universe
$\sim 10^{15}~{\rm GeV}$.  Recent analysis gives the constraint on the
extra radiation component as $N_{\rm eff} = 3.71^{+0.49}_{-0.45}$ at
1$\sigma$ C.L.  \cite{Steigman:2012ve}.  We adopt the 95~\% C.L. upper
limit of $N_{\rm eff}^{\rm (upper)} = 4.65$ in the following analysis.
In the appendix, we provide analytic estimate of the upper bounds on
$n_T$, and show its dependence on the reheating temperature $T_R$.

\subsection{Results}
\begin{figure}[htbp]
  \begin{center}
    \resizebox{160mm}{!}{
     \includegraphics{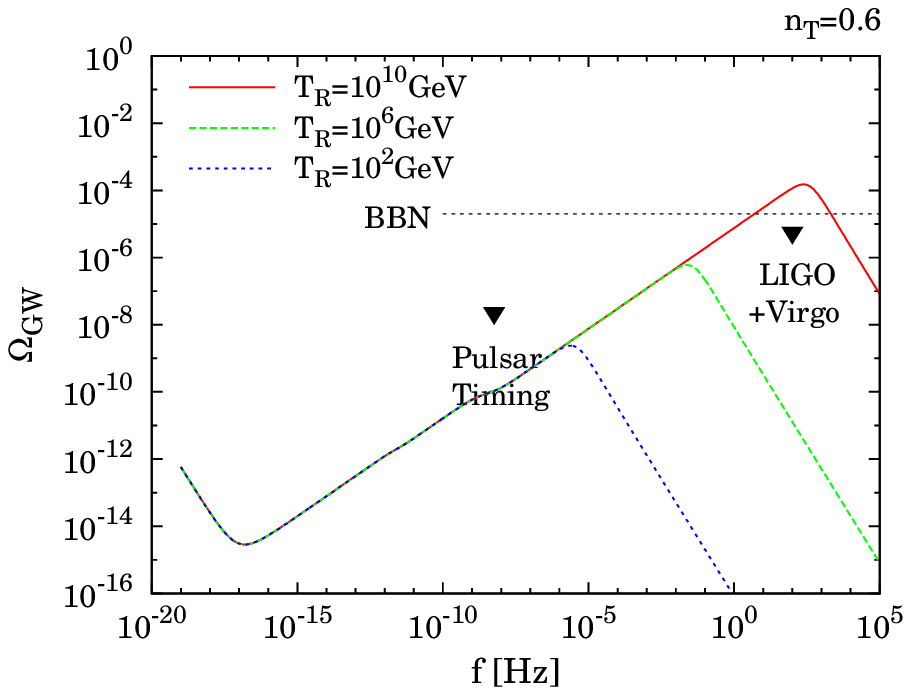}
     \includegraphics{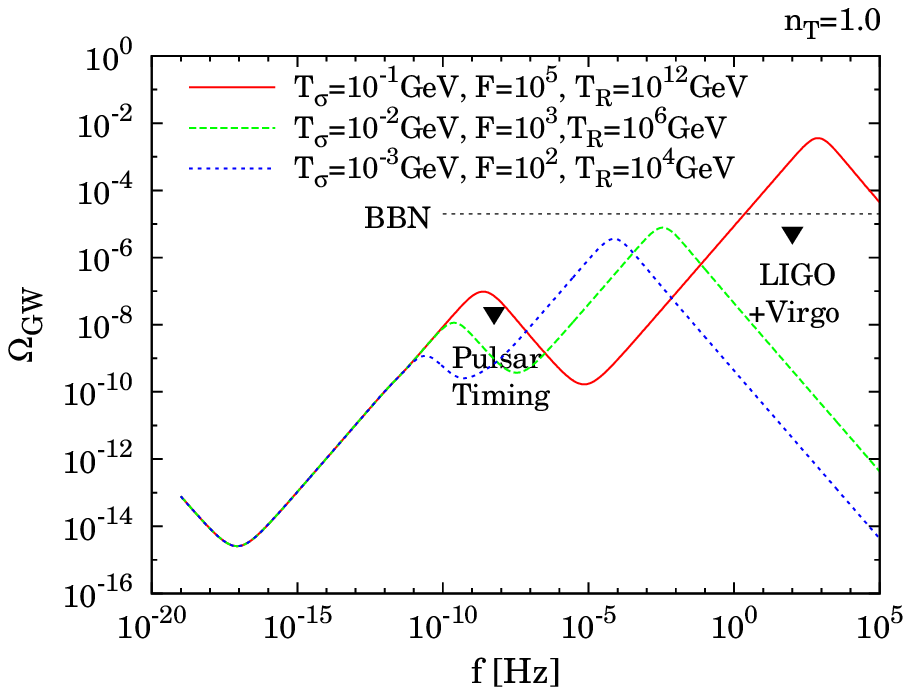}
   }
  \end{center}
  \caption{The GW spectra for the cases with the standard thermal
    history (left) and entropy production due to the decay of the
    $\sigma$-field (right).  In the left panel, spectra with $n_T=0.6$
    are shown by changing reheating temperature $T_R$.  In the right
    panel, spectra with $n_T=1.0$ are shown for different values of
    $T_R$, $T_\sigma$ and $F$.  In both figures, we assume $r=0.2$. }
  \label{fig:spectrum}
\end{figure}

First, to illustrate how the above observational bounds on
$\Omega_{\rm GW}(f)$ help to provide upper bounds on $n_T$ in
consideration of the reheating and late-time entropy production, in
Fig.~\ref{fig:spectrum}, we show spectra of the stochastic GW
background for different thermal history together with the
observational bounds.  As seen from the figure, large values of $n_T$
can elude the observational constraints if the reheating temperature
is low or the amount of entropy produced at late time is large.

In the following, we investigate constraints on $n_T$ for the cases of
the standard reheating and late-time entropy production scenarios, and
show how the constraint depends on parameters characterizing the
thermal history such as $T_R, T_\sigma$ and $F$.  We fix the
tensor-to-scalar ratio at the reference scale $k_{\rm ref} = 0.01 {\rm
  Mpc}^{-1}$ to be $r = 0.2$ in the following analysis.

\subsubsection{Case with the standard reheating scenario}
\label{sec:standard}
We first consider the case of the standard reheating scenario.  In
this scenario, the Universe enters the MD-like epoch soon after the
end of inflation, and is connected to the RD phase when the
temperature of the Universe becomes $T=T_R$. In some models, the
thermal history may be different from the standard one assumed here,
but in such a case, we can apply constraints for a more general case
which will be discussed in Sec.~\ref{sec:general}.  For given $n_T$
and $T_R$, we calculate the stochastic GW spectrum with the method
described in Sec.~\ref{sec:spectrum} and compare it with the current
observational bounds, Eqs.~\eqref{eq:LIGO}, \eqref{eq:NANOGrav} and
\eqref{eq:BBN}.

\begin{figure}[htbp]
  \begin{center}
    \resizebox{100mm}{!}{
     \includegraphics[width=0.5\textwidth]{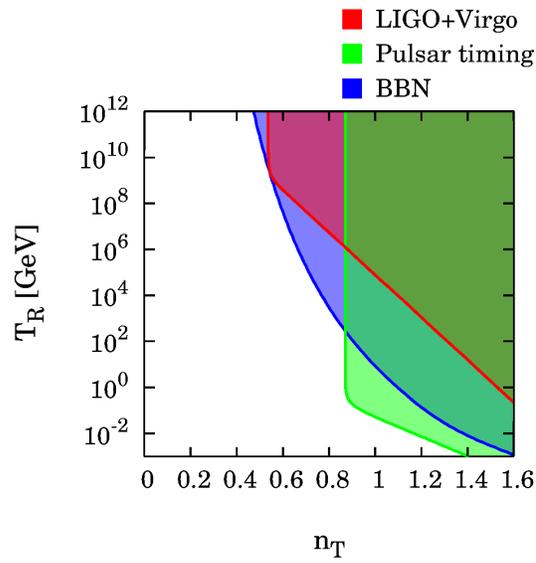}
   }
  \end{center}
  \caption{2$\sigma$ excluded region (colored) in the $n_T$--$T_R$
    plane for the case of the standard reheating scenario.  The
    tensor-to-scalar ratio is assumed to be $r=0.2$. }
  \label{fig:nt_TR}
\end{figure}

In Fig.~\ref{fig:nt_TR}, we show the parameter space ruled out by
LIGO, pulsar timing and BBN in the $n_T-T_R$ plane.  As for the
constraints from the LIGO and pulsar timing, we find a characteristic
temperature above which the constraint on $n_T$ does not depend on the
reheating temperature.  This is because, the reheating temperature
characterizes the frequency of the suppression due to reheating as
seen in Fig.~\ref{fig:spectrum}, and determine the peak frequency of
the spectrum (for $0<n_T<2$).  For a reheating temperature higher than
a certain value, the suppression occurs at the frequencies higher than
the frequency band of the experiment, and the observational bound on
$n_T$ is determined regardless of the effect of reheating.  In
contrast, for a lower reheating temperature, the effect of reheating
becomes important at the frequency of the experiment, and the
constraint on $n_T$ is weakened for smaller $T_R$.  The BBN can put a
relatively severe constraint on $n_T$ depending on the reheating
temperature.  This is because the bound is subject to the integrated
value given in Eq.~\eqref{eq:BBN}.  By putting together all the
constraints, we see the tendency that the constraint on $n_T$ is
relaxed for lower reheating temperatures.  In order not to spoil the
success of BBN, demanding that the reheating temperature should be
larger than about $10~{\rm MeV}$, which indicates that the spectral
index $n_T$ should be smaller than 1.2 for $r = 0.2$.

\subsubsection{Case with late-time entropy production scenario}
Next, we consider the case with late-time entropy production scenario. 
If large entropy is produced by the decay of another scalar field
$\sigma$, the GW spectrum is further suppressed as illustrated in the
right panel of Fig.~\ref{fig:spectrum}.  The degree of the suppression
depends on the amount of the entropy produced, which is characterized
by the quantity $F$ defined in Eq.~\eqref{eq:def_F}.  The frequency
range of the suppression depends on the temperature of the Universe at
the end of the second reheating $T_\sigma$.  Therefore, the bound on
$n_T$ depends on both $F$ and $T_\sigma$.  While it also depends on
the reheating temperature $T_R$ as seen in Fig.~\ref{fig:spectrum},
here we focus on the parameters characterizing the late-time entropy
production, $F$ and $T_\sigma$.  Hence, we fix the reheating
temperature $T_R$ to be rather high as $T_R = 10^{15}~{\rm GeV}$ in
order not to include the effect of the first reheating in the constraints on
$n_T$ and to simply see the tendency of the effect of the late-time
entropy production.

\begin{figure}[htbp]
  \begin{center}
     \includegraphics[width=0.49\textwidth]{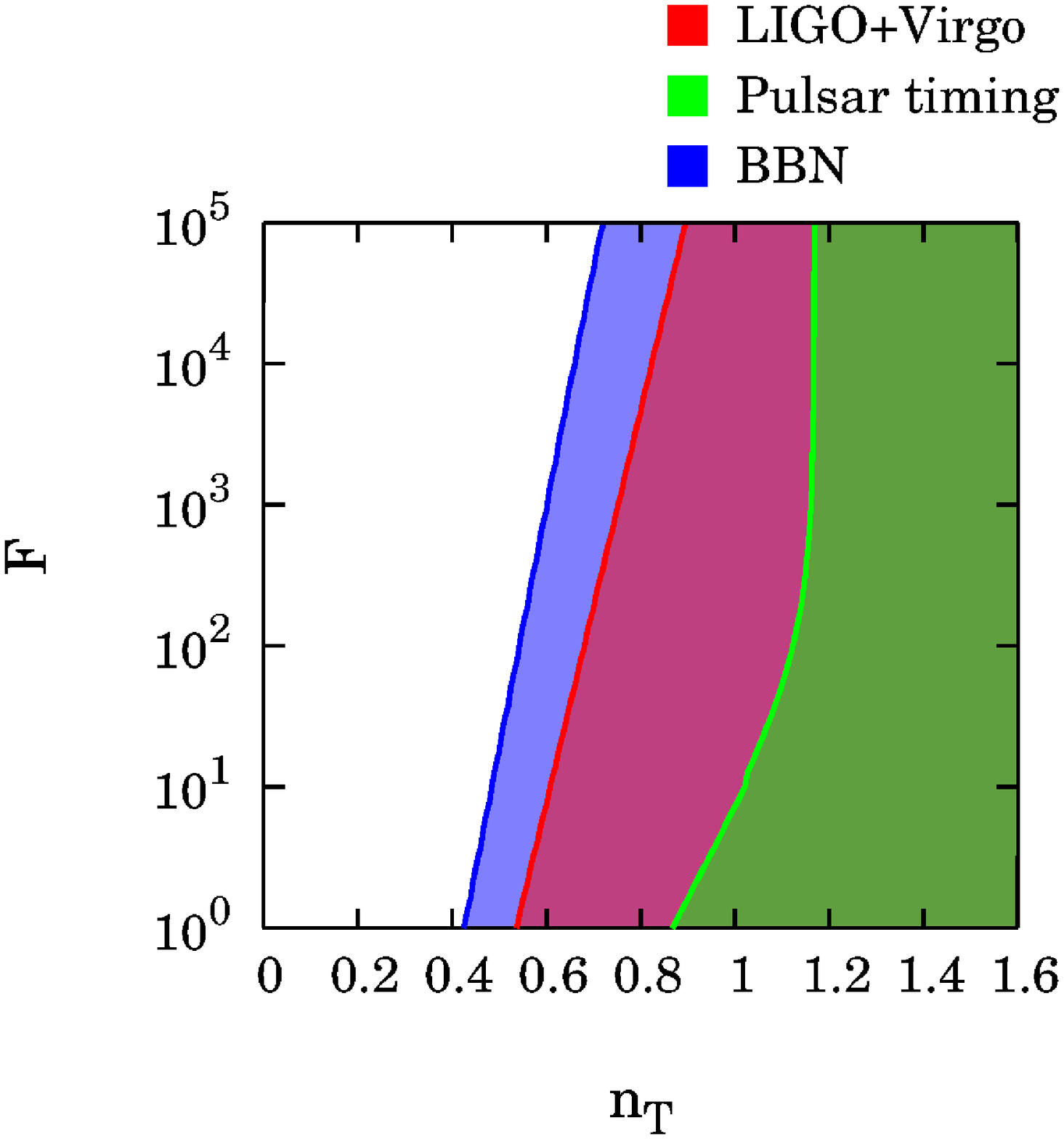}
     \includegraphics[width=0.49\textwidth]{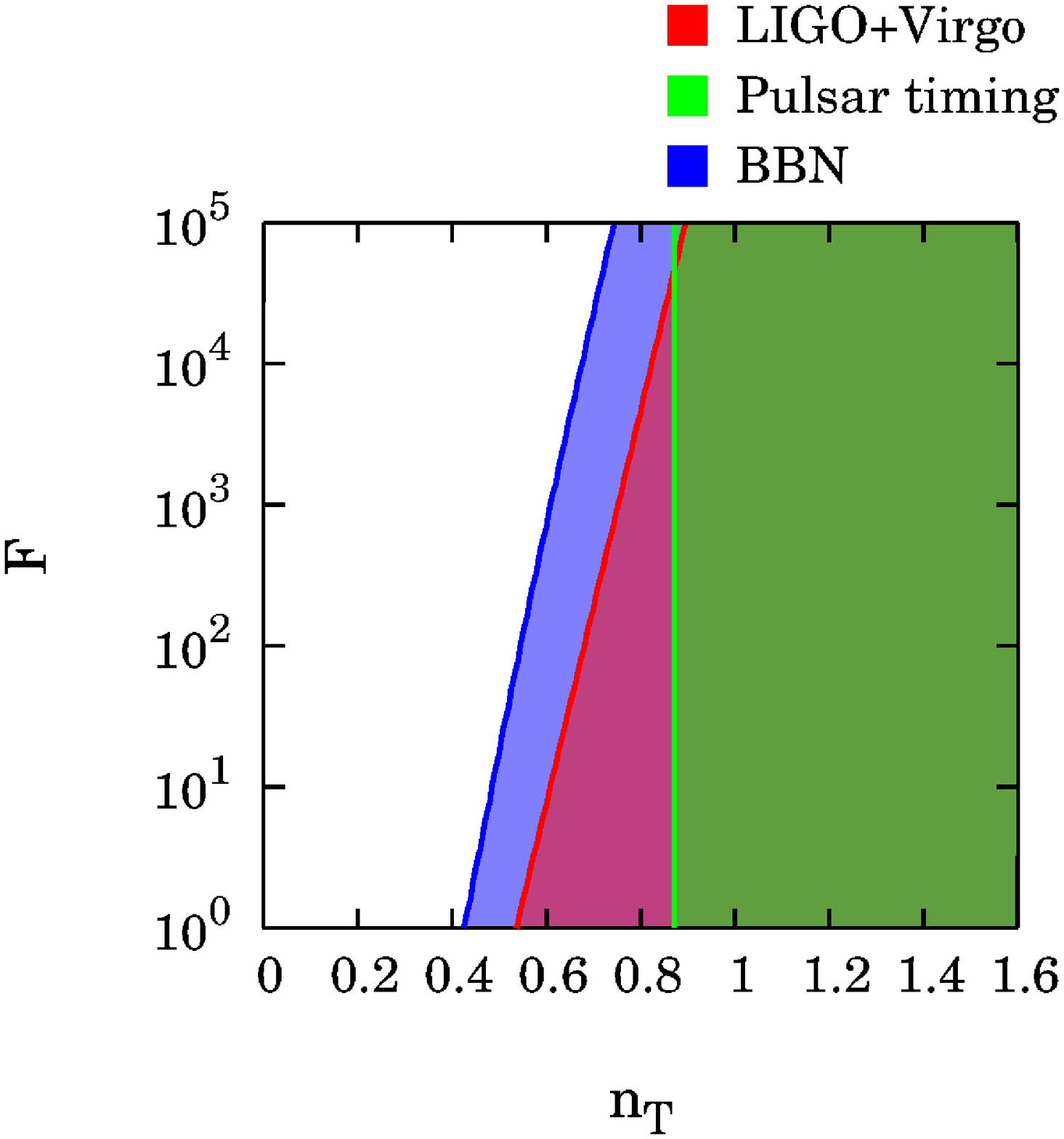}
  \end{center}
  \caption{2$\sigma$ excluded region (colored) in the $n_T$--$F$ plane
    for the cases of $T_\sigma= 10^{-2}$ GeV (left) and $10^3$ GeV
    (right).  We assume $r=0.2$ and $T_R=10^{15}$~GeV.}
  \label{fig:nt_F}
\end{figure}
\begin{figure}[htbp]
  \begin{center}
     \includegraphics[width=0.49\textwidth]{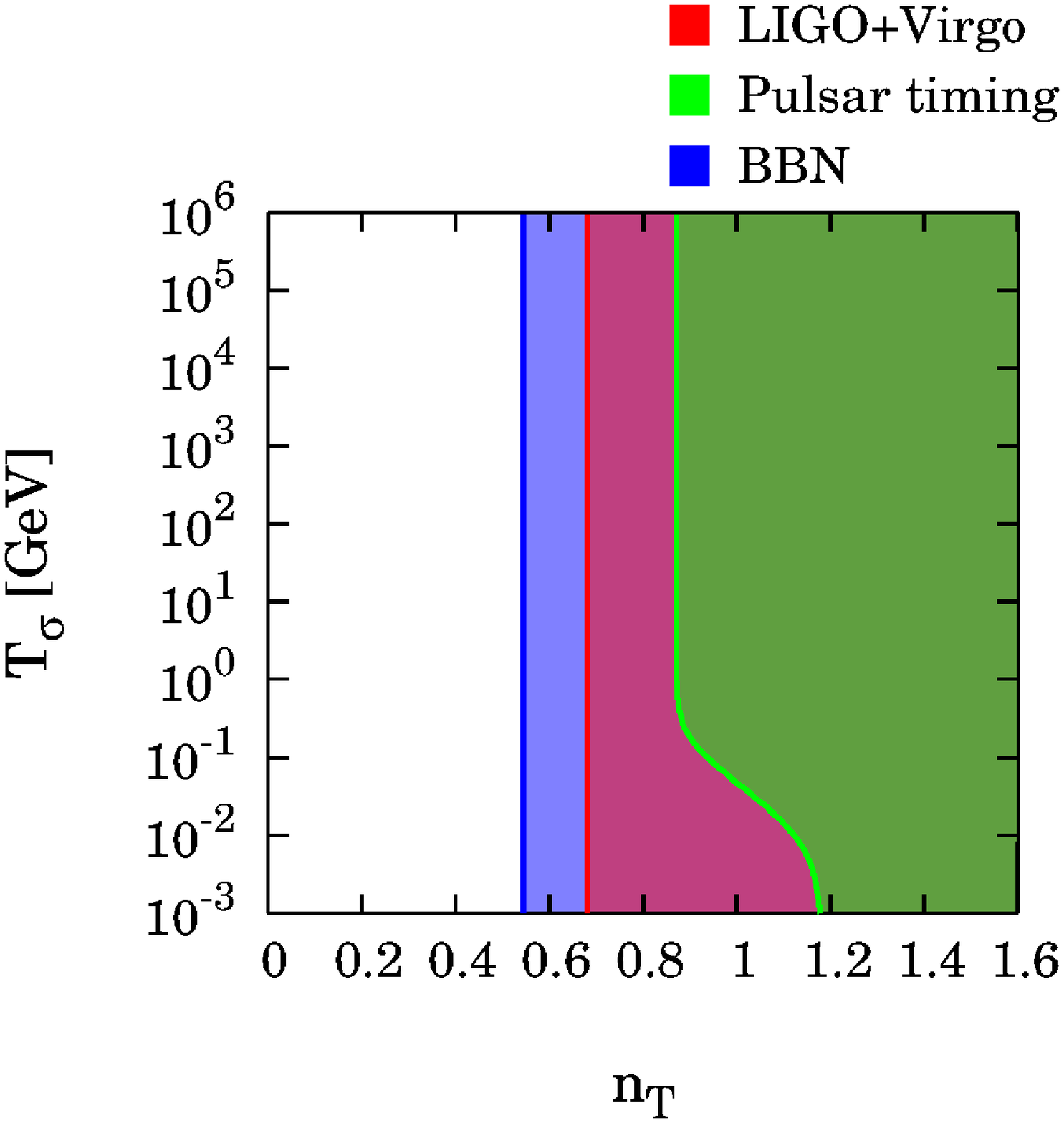}
     \includegraphics[width=0.49\textwidth]{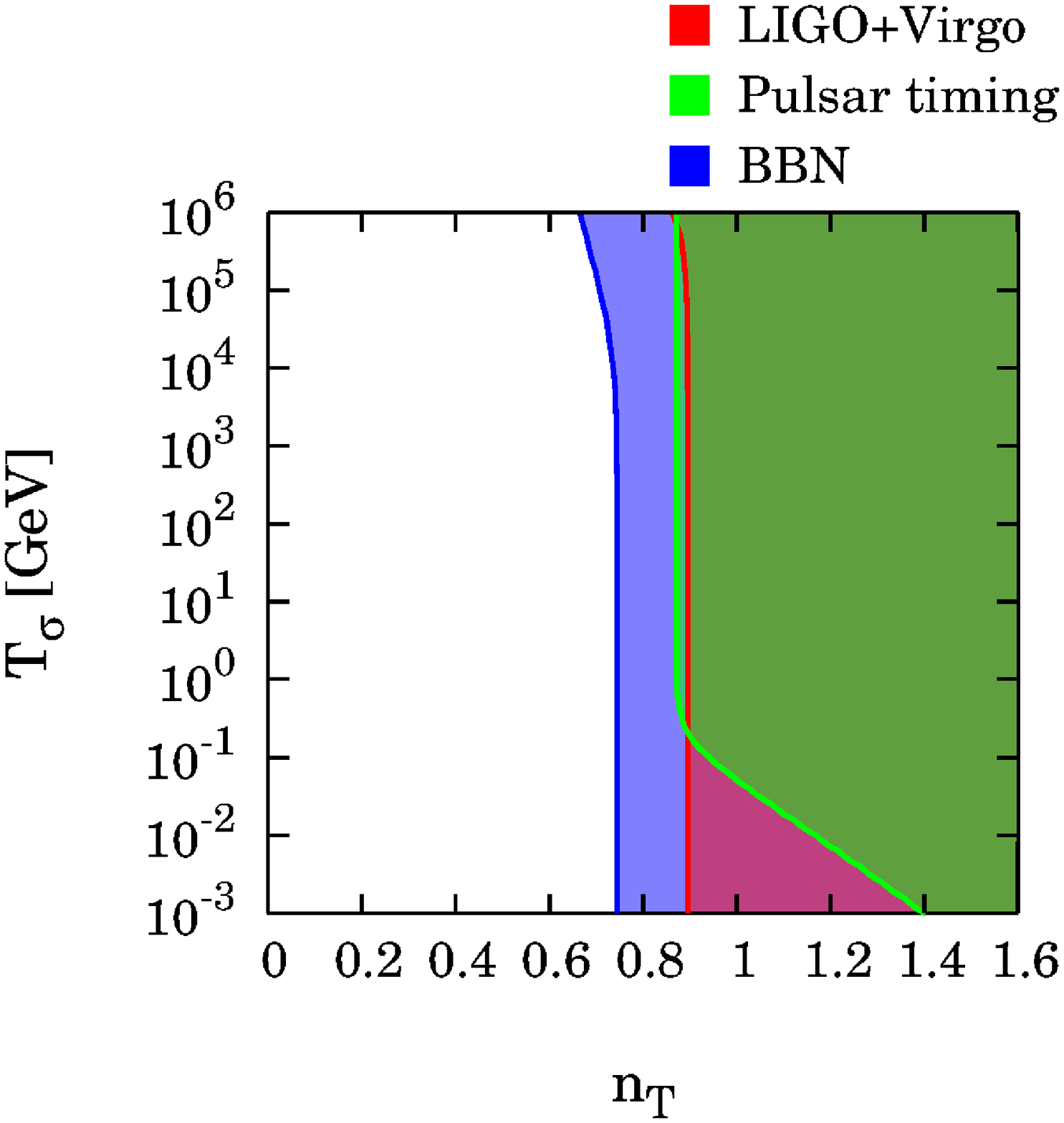}
  \end{center}
  \caption{2$\sigma$ excluded region (colored) in the
    $n_T$--$T_\sigma$ plane for the cases of $F=10^2$ (left) and
    $10^5$ (right).  We assume $r=0.2$ and $T_R=10^{15}$~GeV.}
  \label{fig:nt_Tsig}
\end{figure}

To see the parameter dependence of the constraint, we show excluded
parameter space in the $n_T$--$F$ and $n_T$--$T_\sigma$ planes in
Figs.~\ref{fig:nt_F} and \ref{fig:nt_Tsig}, respectively.  First, in
Fig.~\ref{fig:nt_F}, $T_\sigma$ is fixed to be $10^{-2}$GeV (left
panel) and $10^3$ GeV (right panel).  Since a larger value of $F$
gives a larger suppression of the GW spectrum, constraints from LIGO
and BBN on $n_T$ are weakened as $F$ becomes large.  On the other
hand, especially in the right panel, we see the upper bound on $n_T$
obtained from pulsar timing does not depend on the value of $F$.  This
is because the constraint from pulsar timing is put at $f \sim
10^{-8}~{\rm Hz}$, which corresponds to the mode who entered the
horizon at $T \sim 1~{\rm GeV}$.  Thus, for the case of $T_\sigma>
1~{\rm GeV}$, the spectrum is suppressed at frequencies higher than $f
\sim 10^{-8}~{\rm Hz}$ and thus the constraint from pulsar timing is
irrelevant to the value of $F$.

In Fig. \ref{fig:nt_Tsig}, we show the constrains in the
$n_T$--$T_\sigma$ plane by fixing the value of $F$ to be $10^2$ and
$10^5$.  Since $T_\sigma$ determines the frequency of the suppression,
the tendency is expected to be similar to that of $T_R$ obtained in
Sec.~\ref{sec:standard} for the standard reheating scenario.  However,
in contrast to the case of the standard reheating scenario, we find
that the constraints on $n_T$ from LIGO and BBN do not change
depending on the value of $T_\sigma$.  The reason is that the
suppression due to late-time entropy production arises in a certain
range of frequency, which is determined by the amount of entropy
production~$F$, and the spectrum becomes blue again at high frequency.
The values $F=10^2$ and $10^5$ assumed here are not enough to reduce
the GW amplitude at high frequency and to relax the observational
bound.  Note that since we assume $T_R=10^{15}$GeV, the blue-tilted
spectrum continues to rise towards high frequencies.  This results in
the strong constraint from BBN, since the BBN bound is valid even at
very high frequency.

\begin{figure}[htbp]
  \begin{center}
     \includegraphics[width=0.49\textwidth]{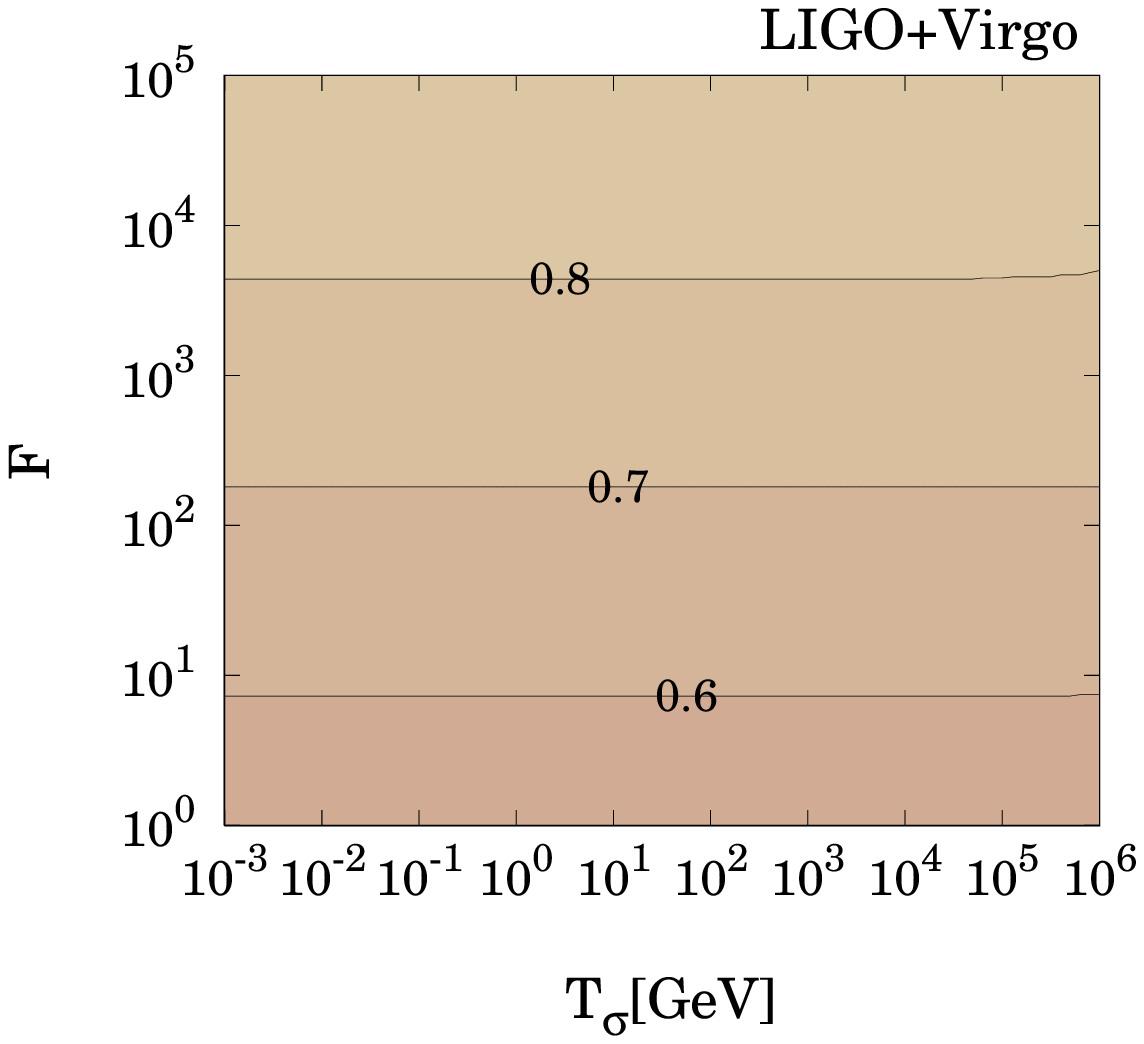}
     \includegraphics[width=0.49\textwidth]{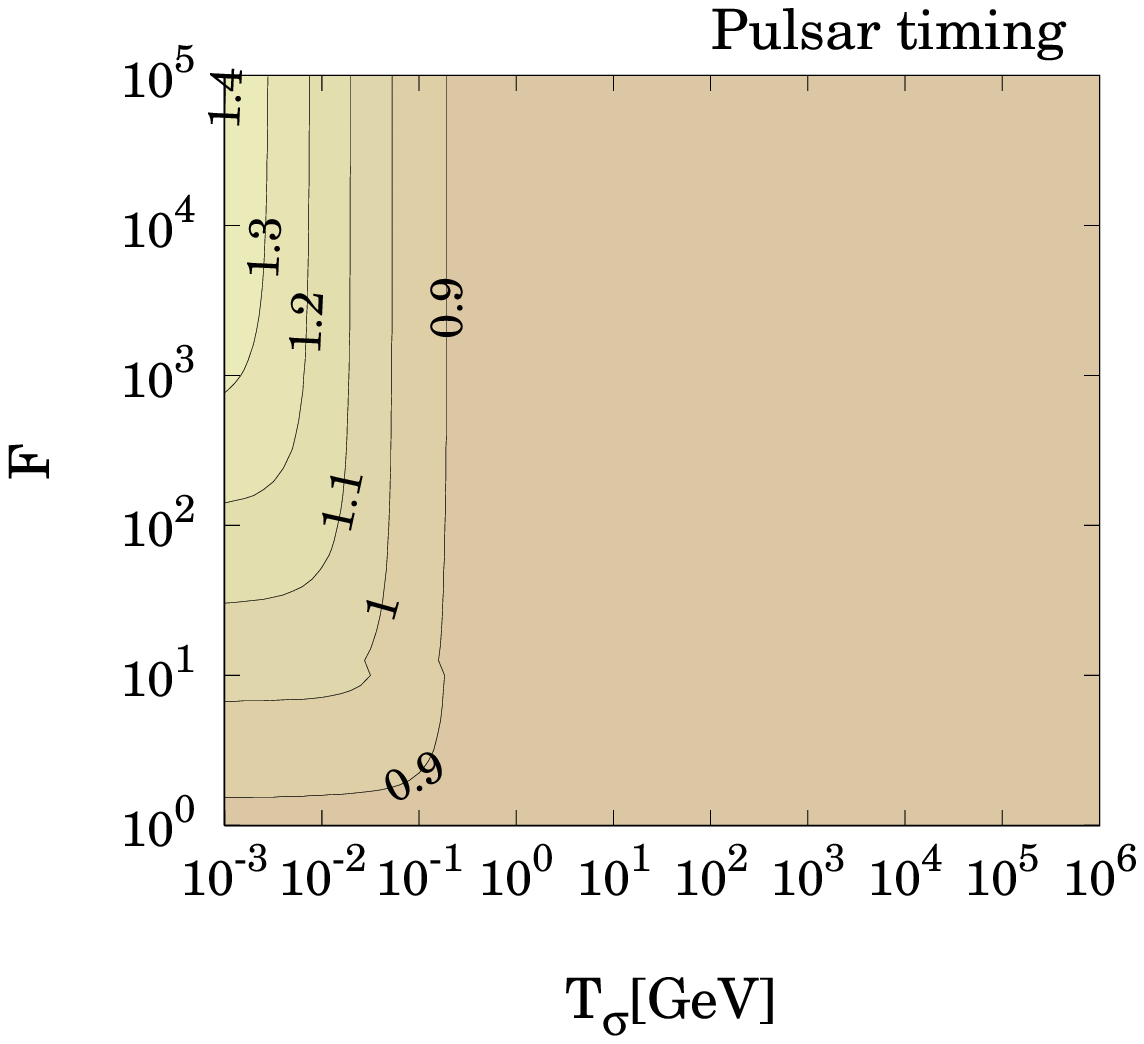}
     \includegraphics[width=0.49\textwidth]{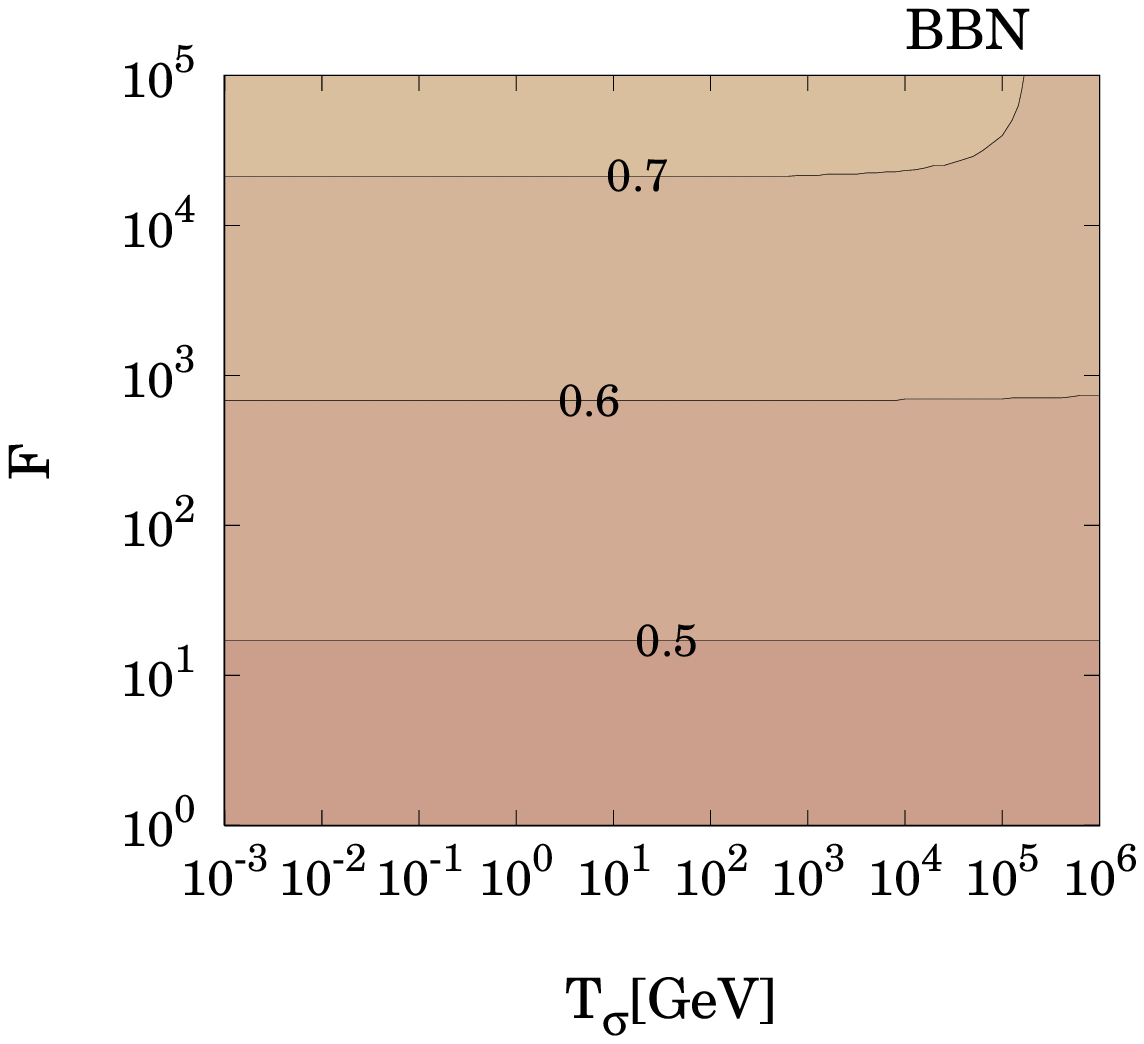}
  \end{center}
  \caption{\label{fig:Tsig_F} Contours of the upper bound on $n_T$ in
    the $T_\sigma$--$F$ plane obtained from LIGO (top left), pulsar
    timing (top right) and BBN (bottom).  Here, $r=0.2$ and
    $T_R=10^{15}$ GeV is assumed.}
\end{figure}

As we have seen in Figs.~\ref{fig:nt_F} and \ref{fig:nt_Tsig}, once
the parameters $T_\sigma, F$, $T_R$ and $r$ are fixed, we can obtain
an upper bound on $n_T$.  In Fig.~\ref{fig:Tsig_F}, we show contours
of the 2$\sigma$ upper bound value of $n_T$ in the $T_\sigma$--$F$
plane for each observational constraint, by fixing the other
parameters as $T_R=10^{15}$~GeV and $r=0.2$. Since LIGO is sensitive at
high frequency $f\sim 100$~Hz, as far as $T_\sigma < 10^{10}$~GeV, the
upper bound on $n_T$ does not depend on $T_\sigma$, and only be
affected by the value of $F$.  On the other hand, the pulsar timing
has sensitivity at relatively low frequency $f\sim 10^{-8}$~Hz.  When
$T_\sigma$ is below $\mathcal{O}(0.1)$~GeV, the suppression becomes
important at $f\sim 10^{-8}$~Hz, and we see the upper bound on $n_T$
is indeed weakened.  From the figure, we can see that BBN provides a
stringent upper bound on $n_T$ in most parameter space.  One of the
reason is that, the reheating temperature is assumed to be very high
$T_R=10^{15}$~GeV and the spectrum is not suppressed by reheating,
which enables the BBN to put strong constraint at very high frequency
region.  If we take a lower value for $T_R$, the BBN constraint
becomes less stringent.

\subsection{Extension to more general case} \label{sec:general} 
In the last subsection, we study the more general case where the stochastic
GW spectrum is modeled by two parameters $\alpha$ and $f_\alpha$, such
that the power index of the spectrum changes from $n_T$ to $\alpha$ at
a characteristic frequency $f_\alpha$.  This modeling covers a broad
class of models of the early Universe, and enables us to provide more
general discussions.  For example, one can interpret the change of the
power-law behavior as an effect of the change of the background
evolution, which is applicable not only for reheating but also for
different models (e.g. kination dominated phase).  One can also adjust
parameter values to describe particular generation mechanism of
primordial GWs which does not predict a uniform spectral index over
whole frequency range.

Here we model the GW spectrum as
\begin{equation}
\label{eq:general_PT}
\Omega_{\rm GW} (k) = 
 \left\{ \begin{array}{ll}
\displaystyle\frac{1}{12} \left( \displaystyle\frac{k}{aH} \right)^2 T_T^2 (k) 
A_T (k_{\rm ref})  \left( \displaystyle\frac{k}{k_{\rm ref}} \right)^{n_T} 
\qquad (k < k_\alpha), \\
\notag \\
\displaystyle\frac{1}{12} \left( \displaystyle\frac{k}{aH} \right)^2 T_T^2 (k) 
A_T (k_{\rm ref})  \left( \displaystyle\frac{k_\alpha}{k_{\rm ref}} \right)^{n_T}
 \left( \displaystyle\frac{k}{k_\alpha} \right)^{\alpha}
\qquad (k > k_\alpha), \\
\end{array}
 \right.
\end{equation} 
where $k_\alpha$ is the frequency at which the power index of the
spectrum changes from $n_T$ to $\alpha$.  The corresponding frequency
$f_\alpha$ is given by $f_\alpha=k_\alpha/2\pi$.  Here, we assume that
$T_T^2(k)$ includes only $T_1^2(k)$, not $T_2^2(k)$ and $T_3^2(k)$,
which means that the change of the frequency dependence due to
matter-radiation equality is included, but the effect of reheating and
late-time entropy production is not included in the transfer function.

The advantage of this modeling is that it can accommodate the change
of the expansion rate before the BBN epoch in a general way.  For
example, we can reproduce the spectrum for the standard reheating
scenario by choosing $\alpha$ to be $-2+n_T$ and $f_\alpha$ to be the
same as $f_R=k_R/2\pi$, given in Eq.~(\ref{eq:k_R}).~\footnote{
  The transfer functions $T_2^2(k)$ and $T_3^2(k)$ are obtained by
  taking into account the decay of the inflaton by following the
  equation of motion as well as the evolution of the Hubble parameter.
  Hence, the GW spectrum given by this modeling does not describe the
  smooth transition in the same way as the transfer functions do, but
  still useful as an approximated form of the spectrum.
}In the case where the kinetic energy of the scalar field dominates
the Universe, it is known that the expansion rate evolves as
$H^2\propto a^{-6}$ and the spectrum of GW has a frequency dependence
of $\Omega_{\rm GW}\propto f$ \cite{Tashiro:2003qp}.  Generally,
assuming that the Universe is dominated by a fluid whose equation of
state is $w$.  GWs which enter the horizon during that phase have a
spectral power dependence of
\begin{equation}
\Omega_{\rm GW} \propto k^{\frac{2(3w-1)}{1 + 3w}}.
\end{equation}
Therefore, when the background equation of state changes from $w$ to
$1/3$ (radiation) at the temperature $T_\alpha$, one can describe the
GW spectrum by assuming $\alpha$ as
\begin{equation}
\alpha = \frac{2(3w-1)}{1 + 3w} + n_T,
\end{equation}
and $f_\alpha$ would be the quantity determined by $T_\alpha$.
 
In addition, in some models predicting a blue-tilted GW spectrum, the
generation mechanism of the primordial GWs does not continuously
generate a blue-tilted spectrum, and the spectral index changes from
blue-tilted one to another.  The general modeling is useful to
describe such change of the spectral power dependence in the
primordial spectrum $\mathcal{P}^{\rm prim}_T (k)$.  For example, if
there exists some phase during inflation in which the production of
the vector quanta becomes effective, for the modes which exit the
horizon during such phase, the generated primordial GWs can be
blue-tilted \cite{Cook:2011hg,Mukohyama:2014gba}.  This model is based
on the action such as ${\psi \over 4 f} F_{\mu\nu} \tilde{F}^{\mu\nu}$
where $F_{\mu\nu}$ and $\tilde{F}^{\mu\nu} =
\epsilon^{\mu\nu\alpha\beta}F_{\alpha \beta}$ are the field strength
of the gauge field and its dual, $\psi$ is a dynamical pseudo-scalar
field and $f$ is a pseudo-scalar decay constant.  Through the coupling
between the pseudo-scalar and gauge field, time dependent
pseudo-scalar $\psi$ induces the production of the gauge quanta and
this gauge quanta becomes the source of the gravitational waves.
Since the spectral tilt of the gravitational waves depends on the time
evolution of the pseudo-scalar $\dot{\psi}$, the blue tensor spectrum
can be realized by taking into account appropriate dynamics of $\psi$.
For the modes which cross the horizon after the production phase
(e.g., $\psi$ stops the rolling), the GW spectrum is not blue-tilted
anymore, and the amplitude of the primordial GWs decreases at higher
frequency.  In a such case, the characteristic frequency $f_\alpha$
represents the mode which exits the horizon at the end of the
effective particle creation phase during inflation.  Then, the
duration of the particle creation phase, measured from the time when
the reference scale exit the horizon, can be expressed in terms of the
$e$-folding number as
\begin{eqnarray}
\Delta N_{\rm p.c.} \equiv \ln \left( {a_\alpha \over a_{\rm ref}}\right) \simeq \ln \left( {k_\alpha \over k_{\rm ref}}\right),
\end{eqnarray}
where we assume that the Hubble parameter during inflation stays
almost constant.

\begin{figure}[htbp]
  \begin{center}
     \includegraphics[width=0.49\textwidth]{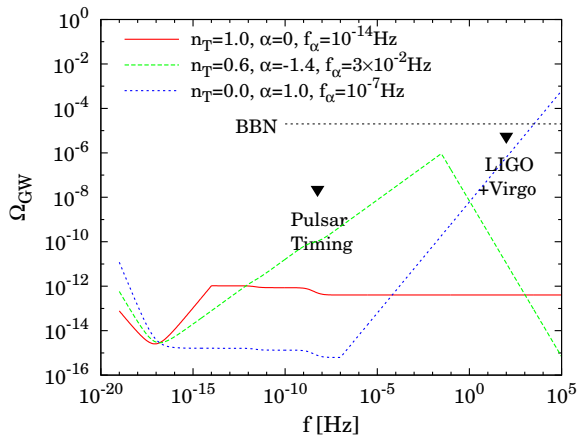}
  \end{center}
  \caption{GW Spectra modeled by $\alpha$ and $f_\alpha$.}
  \label{fig:spectrum_general}
\end{figure}

To obtain an intuitive idea of this modeling and how the constraint on
$n_T$ depends on the value of $\alpha$ and $k_\alpha$ (or $f_\alpha$),
we show the GW spectrum represented by Eq.~\eqref{eq:general_PT} in
Fig.~\ref{fig:spectrum_general}.  In this figure, we show three
examples of GW spectrum considering different scenarios of the early
Universe.  The green dashed spectrum corresponds to a blue-tilted
primordial spectrum, $n_T=0.6$, which underwent the standard reheating
scenario with $T_R=10^6$~GeV.  The blue dotted one describes a
scale-invariant primordial spectrum which has experienced a
kination-dominated phase after inflation.  For the red solid one, we
consider a primordial spectrum such that the model predicts a strongly
blue-tilted spectrum of $n_T=1$ around the CMB scale, but at some
later stage, the situation become the same as the standard slow-roll
inflation.

\begin{figure}[htbp]
  \begin{center}
     \includegraphics[width=0.49\textwidth]{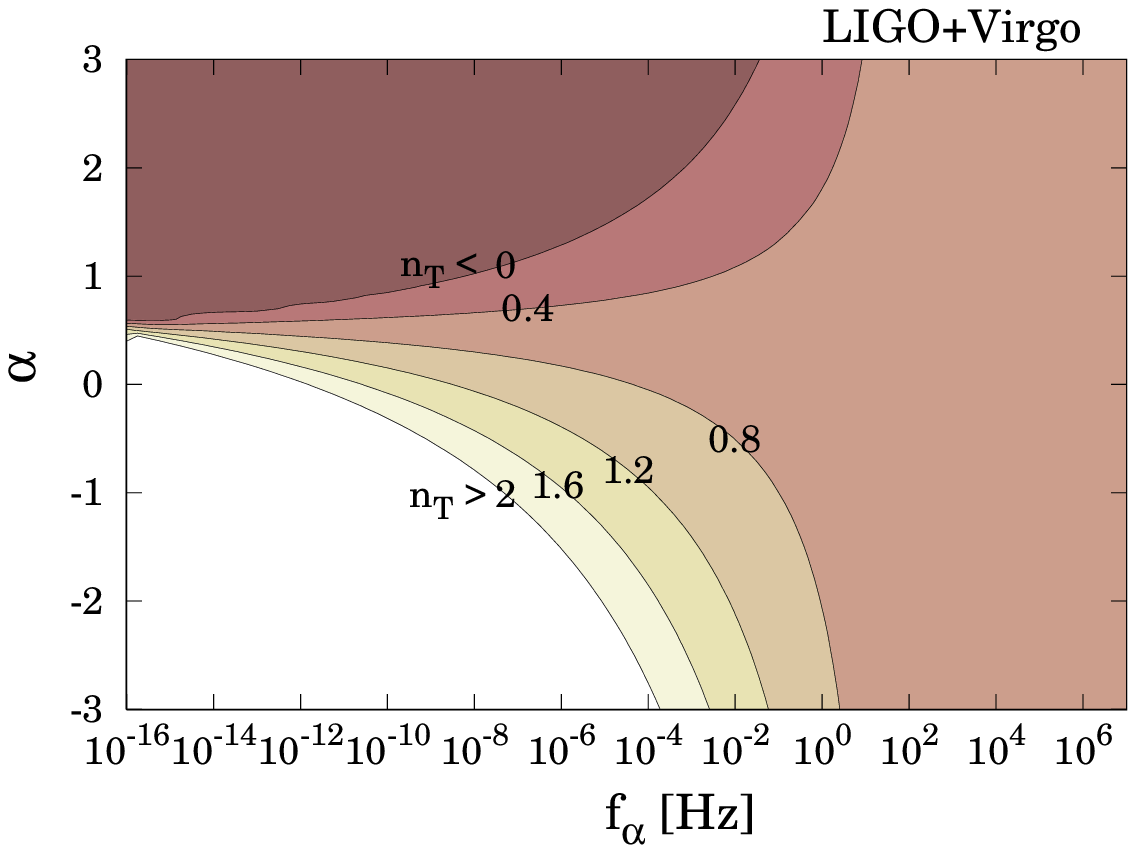}
     \includegraphics[width=0.49\textwidth]{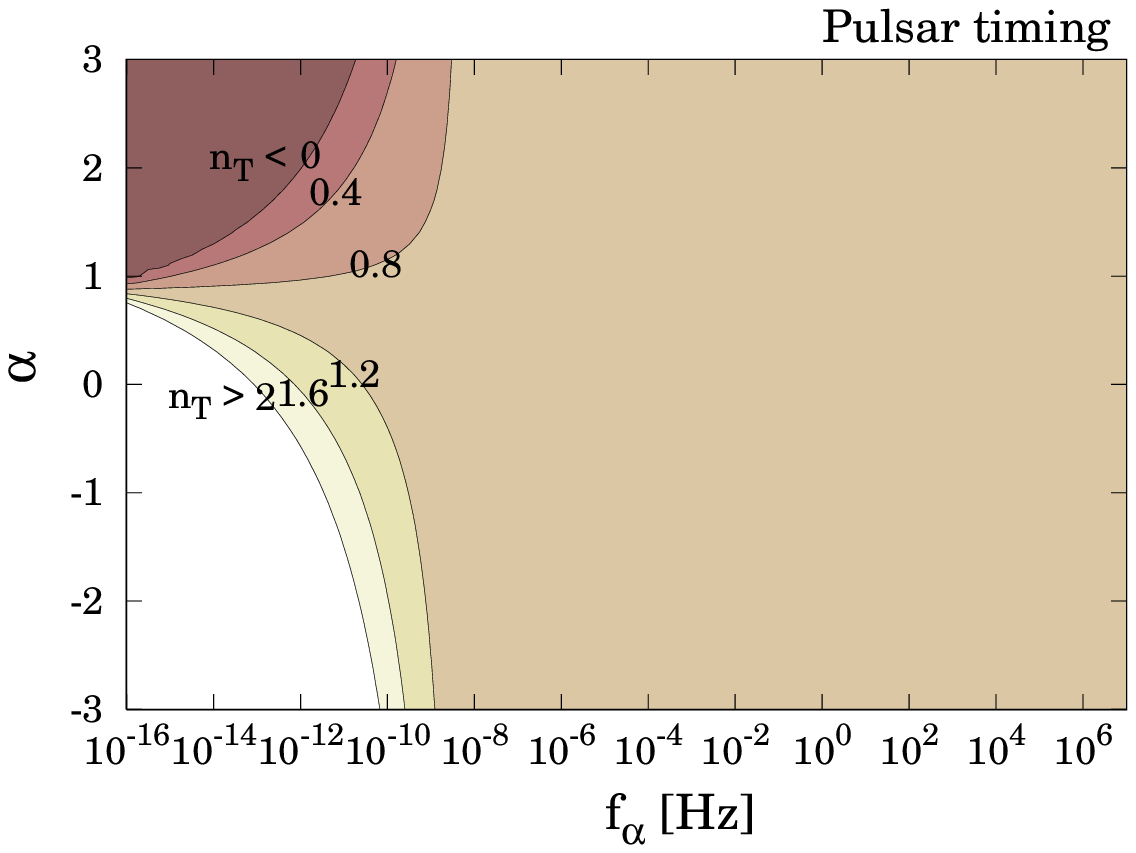}
     \includegraphics[width=0.49\textwidth]{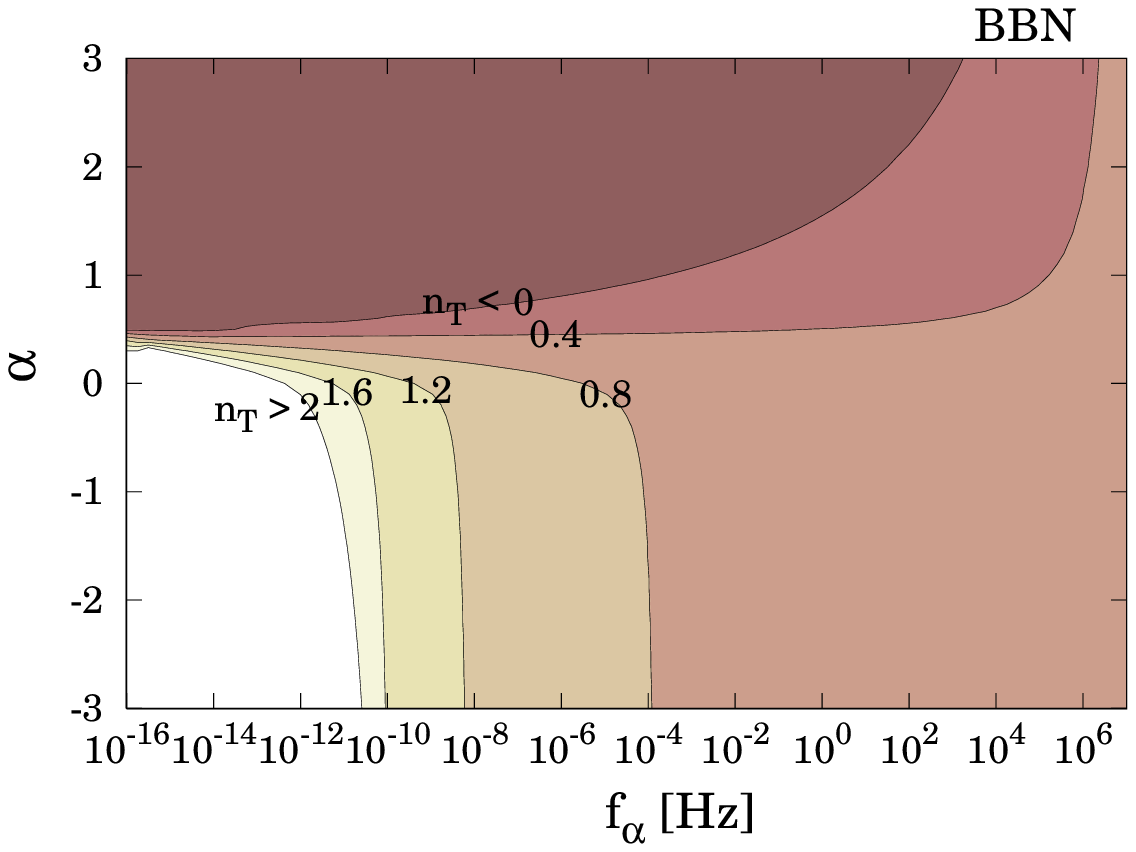}
  \end{center}
  \caption{Contours of the 2$\sigma$ upper bound on $n_T$ in the
    $f_\alpha$--$\alpha$ plane.  Here $r=0.2$ is assumed.}
  \label{fig:general_const}
\end{figure}

Finally, in Fig.~\ref{fig:general_const}, the upper bound on $n_T$ is
shown as a contour plot in the $f_\alpha$--$\alpha$ plane for each
observation, i.e., LIGO, pulsar timing and BBN.  Once we know an
approximated form of the GW spectrum for some particular model, we can
find the values of $\alpha$ and $f_\alpha$ for that model and read off
the upper bound on $n_T$ from the figure.  In the figure, we can see
the tendency that the constraint on $n_T$ becomes tighter for larger
$\alpha$, since a positive value of $\alpha$ enhances the spectral
amplitude at high frequencies.  For negative values of $\alpha$, we can
see the constraint becomes weaker when $f_\alpha$ is smaller than the
frequency of the observational bound.  This is because, for smaller
$f_\alpha$, a negative value of $\alpha$ reduces the spectral amplitude
at the corresponding frequency of the experiment.  On the other hand,
for larger $f_\alpha$, the suppression occurs at higher frequency and
does not affect the spectrum at the frequency where the observational
bounds are relevant.

\section{Conclusion}  \label{sec:conclusion}
We have provided more general constraints on the blue-tilted GW
spectrum using observational bounds from LIGO, pulsar timing and BBN,
taking into account the effect of the thermal history after inflation,
such as reheating and late-time entropy production.  The recent
finding of CMB B-mode polarization by BICEP2 suggests the large
tensor-to-scalar ratio $r \sim 0.2$ and some analysis pointed out that
the data favors a blue-tilted spectrum.  Besides, from the theoretical
point of view, there are several models of the early Universe which
predict a blue-tilted spectrum, hence it is worth investigating to
what extent blue-tilted spectrum can be allowed by current
observational bound on the stochastic GW amplitude.

First, we have found that the suppression of the GW spectrum due to
reheating can significantly relax the constraint on the tensor
spectral index $n_T$, depending on the reheating temperature $T_R$.
We have also shown that, by taking into account the late-time entropy
production, the constraint on $n_T$ changes depending on the amount of
the produced entropy $F$ and the cosmic temperature at the epoch of
entropy production $T_\sigma$.  When one does not take into account
the thermal history (or assuming a very high reheating temperature
such as $T_R = 10^{15}$~GeV) and assume that the spectral amplitude
continue to increase at high frequencies with the spectral index
$n_T$, the constraints on $n_T$ from BBN, LIGO and pulsar timing are
$n_T < 0.43$, $0.54$ and $0.87$ for $r=0.2$ at 95~\% C.L.,
respectively.  These bounds are modified depending on the assumption
of the thermal history: for example, if we consider reheating with
$T_R = 10^6$~GeV, these bound are weakened to $n_T < 0.64$ (BBN),
$0.89$ (LIGO) and $0.87$ (pulsar).  If we consider a late-time entropy
production after reheating with $F=10^2$, $T_\sigma = 10^{-2}$~GeV and
$T_R = 10^{15}$~GeV, the bounds become $n_T < 0.55$ (BBN), $0.68$
(LIGO) and $1.1$ (pulsar).  Note that the results shown in this paper
are all calculated assuming $r=0.2$, motivated by the recent BICEP2
result.  These bound on $n_T$ changes for different values of $r$.
Since $r$ and $n_T$ are degenerate in the expression of the spectrum
as $\Omega_{\rm GW}\propto r(k/k_{\rm ref})^{n_T}$, the constraints on
$n_T$ are relaxed for small value of $r$, while the dependencies on
the thermal history parameters are the same.  The degree of the
relaxation is determined by the value of $k/k_{\rm ref}$, which has a
different value depending on the experiment.

We have also discussed the constraints using the more generalized
modeling of the spectrum, where the spectral power dependence changes
at the characteristic frequency $f_\alpha$ and we allow a general
power-law form for the spectrum for $f > f_\alpha$.  This modeling
accommodates various models of the early Universe.  We have shown that
a significantly blue-tilted spectrum is still allowed in some
parameter region, while there are also regions of parameter space
where the value of $n_T$ is severely constrained.  Thus, the
constraint on $n_T$ strongly depends on the underlying models of the
early Universe.

In this paper, we have demonstrated that a very blue-tilted GW
spectrum can still be allowed in some particular models of the thermal
history.  However, in the future, various space-based interferometer
experiments are planned, such as eLISA
\cite{AmaroSeoane:2012km,AmaroSeoane:2012je}, BBO
\cite{bbo,Harry:2006fi} and DECIGO \cite{Kawamura:2011zz} and so on.
Once these experiments are realized, they will put a severe constraint
on the spectral index $n_T$, which would surely help to obtain new
insight into the physics of the early Universe.

\section*{Acknowledgments}
The authors thank K. Nakayama and J. Yokoyama for helpful discussions. 
The work is supported by Grant-in-Aid for Scientific Research
Nos.~24-2775~(SY) and 23740195~(TT) from the Ministry of Education,
Culture, Sports, Science and Technology in Japan.

\appendix
\section{Analytic estimate of the BBN bound}
\label{sec:ana}
Although we have numerically calculated the integral in the left hand
side of Eq.~\eqref{eq:BBN} to obtain the BBN constraint in
Sec. \ref{sec:standard}, here we show that the upper bounds on $n_T$
can be analytically estimated and given in terms of reheating
temperature $T_R$.

Assuming the standard thermal history after inflation, one can
approximate the spectrum with a simple broken power-law such as
\begin{eqnarray}
\Omega_{\rm GW} (f) = \left\{ \begin{array}{ll}
\Omega_{\rm GW} (f_R) \left( {f \over f_R} \right)^{n_T} & \qquad {\rm for} \quad f_{\rm eq} < f_1 < f < f_R, \\
\Omega_{\rm GW} (f_R) \left( {f \over f_R} \right)^{n_T - 2} & \qquad {\rm for} \quad f_R < f < f_2. \\ 
\end{array}
 \right.
\end{eqnarray}
Then we can perform the integration in Eq. \eqref{eq:BBN} and obtain
\begin{eqnarray}
{\Omega_{\rm GW} (f_R) h^2 \over n_T}
\left[ 1 - \left( {f_1 \over f_R} \right)^{n_T} \right] + {\Omega_{\rm GW} (f_R) h^2 \over 2 - n_T}
\left[ 1 - \left( {f_2 \over f_R} \right)^{n_T - 2} \right] \le 5.6 \times 10^{-6} (N_{\rm eff}^{\rm (upper)} -3) ,
\end{eqnarray}
for $0< n_T < 2$.  Furthermore, in case with $f_1 \ll f_R \ll f_2$, we
have a simple expression as
\begin{eqnarray}
\Omega_{\rm GW} (f_R) h^2 \le {n_T (2 - n_T) \over 2} \times 5.6 \times 10^{-6} (N_{\rm eff}^{\rm (upper)} -3).
\end{eqnarray}
Since one can also write down $\Omega_{\rm GW} (f_R)$ by using $r,
n_T$ and $T_R$, the above relation gives
\begin{equation}
\log \left( \frac{T_R}{10^6~{\rm GeV}} \right)  
< \frac{1}{n_T} \log ( n_T ( 2 - n_T) ) -35.1 + \frac{24.3}{n_T} 
\simeq -35.1 + \frac{24.3}{n_T}, 
\end{equation}
where we have assumed $k_{\rm ref}=0.01~{\rm Mpc}^{-1}$, $g_\ast = 106.75$
and $r= 0.2$.


\begin{thebibliography}{100}


\bibitem{Ade:2014xna} 
  P.~A.~R.~Ade {\it et al.}  [BICEP2 Collaboration],
  arXiv:1403.3985 [astro-ph.CO].
  
\bibitem{Ade:2014gua} 
  P.~A.~R.~Ade {\it et al.}  [BICEP2 Collaboration],
  arXiv:1403.4302 [astro-ph.CO].
  
\bibitem{Liu:2014mpa} 
  H.~Liu, P.~Mertsch and S.~Sarkar,
  arXiv:1404.1899 [astro-ph.CO].

\bibitem{Mortonson:2014bja} 
  M.~J.~Mortonson and U.~Seljak,
  arXiv:1405.5857 [astro-ph.CO].

\bibitem{Flauger:2014qra} 
  R.~Flauger, J.~C.~Hill and D.~N.~Spergel,
  arXiv:1405.7351 [astro-ph.CO].
  
\bibitem{Ade:2013uln} 
  P.~A.~R.~Ade {\it et al.}  [Planck Collaboration],
  arXiv:1303.5082 [astro-ph.CO].


\bibitem{Gong:2014qga} 
  J.~-O.~Gong,
  arXiv:1403.5163 [astro-ph.CO].


\bibitem{Gerbino:2014eqa} 
  M.~Gerbino, A.~Marchini, L.~Pagano, L.~Salvati, E.~Di Valentino and A.~Melchiorri,
  arXiv:1403.5732 [astro-ph.CO].


\bibitem{Wang:2014kqa} 
  Y.~Wang and W.~Xue,
  arXiv:1403.5817 [astro-ph.CO].


\bibitem{Ashoorioon:2014nta} 
  A.~Ashoorioon, K.~Dimopoulos, M.~M.~Sheikh-Jabbari and G.~Shiu,
  arXiv:1403.6099 [hep-th].


\bibitem{Wu:2014qxa} 
  F.~Wu, Y.~Li, Y.~Lu and X.~Chen,
  arXiv:1403.6462 [astro-ph.CO].
  

\bibitem{Brandenberger:2006xi} 
  R.~H.~Brandenberger, A.~Nayeri, S.~P.~Patil and C.~Vafa,
  Phys.\ Rev.\ Lett.\  {\bf 98}, 231302 (2007)
  [hep-th/0604126].


\bibitem{Baldi:2005gk} 
  M.~Baldi, F.~Finelli and S.~Matarrese,
  Phys.\ Rev.\ D {\bf 72}, 083504 (2005)
  [astro-ph/0505552].


\bibitem{Kobayashi:2010cm} 
  T.~Kobayashi, M.~Yamaguchi and J.~'i.~Yokoyama,
  Phys.\ Rev.\ Lett.\  {\bf 105}, 231302 (2010)
  [arXiv:1008.0603 [hep-th]].

\bibitem{Calcagni:2004as} 
  G.~Calcagni and S.~Tsujikawa,
  Phys.\ Rev.\ D {\bf 70}, 103514 (2004)
  [astro-ph/0407543].

\bibitem{Calcagni:2013lya} 
  G.~Calcagni, S.~Kuroyanagi, J.~Ohashi and S.~Tsujikawa,
  JCAP {\bf 1403}, 052 (2014)
  [arXiv:1310.5186 [astro-ph.CO]].

\bibitem{Cook:2011hg} 
  J.~L.~Cook and L.~Sorbo,
  Phys.\ Rev.\ D {\bf 85}, 023534 (2012)
  [Erratum-ibid.\ D {\bf 86}, 069901 (2012)]
  [arXiv:1109.0022 [astro-ph.CO]].

\bibitem{Mukohyama:2014gba} 
  S.~Mukohyama, R.~Namba, M.~Peloso and G.~Shiu,
  arXiv:1405.0346 [astro-ph.CO].


\bibitem{Jenet:2006sv} 
  F.~A.~Jenet, G.~B.~Hobbs, W.~van Straten, R.~N.~Manchester, M.~Bailes, J.~P.~W.~Verbiest, R.~T.~Edwards and A.~W.~Hotan {\it et al.},
  Astrophys.\ J.\  {\bf 653}, 1571 (2006)
  [astro-ph/0609013].


\bibitem{vanHaasteren:2011ni} 
  R.~van Haasteren, Y.~Levin, G.~H.~Janssen, K.~Lazaridis, M.~Kramer, B.~W.~Stappers, G.~Desvignes and M.~B.~Purver {\it et al.},
  Mon.\ Not.\ Roy.\ Astron.\ Soc.\  {\bf 414}, no. 4, 3117 (2011)
  [Erratum-ibid.\  {\bf 425}, no. 2, 1597 (2012)]
  [arXiv:1103.0576 [astro-ph.CO]].

\bibitem{Demorest:2012bv} 
  P.~B.~Demorest, R.~D.~Ferdman, M.~E.~Gonzalez, D.~Nice, S.~Ransom, I.~H.~Stairs, Z.~Arzoumanian and A.~Brazier {\it et al.},
  Astrophys.\ J.\  {\bf 762}, 94 (2013)
  [arXiv:1201.6641 [astro-ph.CO]].


\bibitem{Zhao:2013bba} 
  W.~Zhao, Y.~Zhang, X.~-P.~You and Z.~-H.~Zhu,
  Phys.\ Rev.\ D {\bf 87}, 124012 (2013)
  [arXiv:1303.6718 [astro-ph.CO]].

\bibitem{Allen:1996vm} 
  B.~Allen,
  In *Les Houches 1995, Relativistic gravitation and gravitational radiation* 373-417
  [gr-qc/9604033].


\bibitem{Maggiore:1999vm} 
  M.~Maggiore,
  Phys.\ Rept.\  {\bf 331}, 283 (2000)
  [gr-qc/9909001].

\bibitem{Aasi:2014zwg} 
  J.~Aasi {\it et al.}  [LIGO Scientific and VIRGO Collaborations],
  arXiv:1406.4556 [gr-qc].

  
\bibitem{Abadie:2011fx} 
  J.~Abadie {\it et al.}  [LIGO Scientific and Virgo Collaborations],
  Phys.\ Rev.\ D {\bf 85}, 122001 (2012)
  [arXiv:1112.5004 [gr-qc]].

\bibitem{Seto:2005qy} 
  N.~Seto,
  Phys.\ Rev.\ D {\bf 73}, 063001 (2006)
  [gr-qc/0510067].

\bibitem{Smith:2006nka} 
  T.~L.~Smith, E.~Pierpaoli and M.~Kamionkowski,
  Phys.\ Rev.\ Lett.\  {\bf 97}, 021301 (2006)
  [astro-ph/0603144].
  
\bibitem{Sendra:2012wh} 
  I.~Sendra and T.~L.~Smith,
  Phys.\ Rev.\ D {\bf 85}, 123002 (2012)
  [arXiv:1203.4232 [astro-ph.CO]].
  

\bibitem{Ota:2014hha} 
  A.~Ota, T.~Takahashi, H.~Tashiro and M.~Yamaguchi,
  arXiv:1406.0451 [astro-ph.CO].
  
\bibitem{Chluba:2014qia} 
  J.~Chluba, L.~Dai, D.~Grin, M.~Amin and M.~Kamionkowski,
  arXiv:1407.3653 [astro-ph.CO].


\bibitem{Siegel:2014yta} 
  D.~M.~Siegel and M.~Roth,
  Astrophys.\ J.\  {\bf 784}, 88 (2014)
  [arXiv:1401.6888 [gr-qc]].


\bibitem{Armstrong:2003ay} 
  J.~W.~Armstrong, L.~Iess, P.~Tortora and B.~Bertotti,
  Astrophys.\ J.\  {\bf 599}, 806 (2003).
  
\bibitem{Hui:2012yp} 
  L.~Hui, S.~T.~McWilliams and I-S.~Yang,
  Phys.\ Rev.\ D {\bf 87}, 084009 (2013)
  [arXiv:1212.2623 [gr-qc]].
  
  
\bibitem{Shoda:2013oya} 
  A.~Shoda, M.~Ando, K.~Ishidoshiro, K.~Okada, W.~Kokuyama, Y.~Aso and K.~Tsubono,
  Phys.\ Rev.\ D {\bf 89}, 027101 (2014)
  [arXiv:1311.4273 [gr-qc]].  


\bibitem{Coughlin:2014sca} 
  M.~Coughlin and J.~Harms,
  Phys.\ Rev.\ Lett.\  {\bf 112}, 101102 (2014)
  [arXiv:1401.3028 [gr-qc]].
  
\bibitem{Akutsu:2008qv} 
  T.~Akutsu, S.~Kawamura, A.~Nishizawa, K.~Arai, K.~Yamamoto, D.~Tatsumi, S.~Nagano and E.~Nishida {\it et al.},
  Phys.\ Rev.\ Lett.\  {\bf 101}, 101101 (2008)
  [arXiv:0803.4094 [gr-qc]].
  
  \bibitem{Astone:1999aa}
  P.~Astone, M.~Bassan, P.~Bonifazi, P.~Carelli, C.~Cosmelli, E.~Coccia, V.~Fafone and S.~Frasca  {\it et al.},
  Astron.\ Astrophys. {\bf 351}, 811 (1999)
  
\bibitem{Aoyama:2014fea} 
  S.~Aoyama, R.~Tazai and K.~Ichiki,
  arXiv:1402.4521 [gr-qc].
  
  
\bibitem{Stewart:2007fu} 
  A.~Stewart and R.~Brandenberger,
  JCAP {\bf 0808}, 012 (2008)
  [arXiv:0711.4602 [astro-ph]].
 
\bibitem{Camerini:2008mj} 
  R.~Camerini, R.~Durrer, A.~Melchiorri, A.~Riotto, 
  Phys.\ Rev.\ D {\bf 77}, 101301 (2008)
  [arXiv:0802.1442 [astro-ph]].
  
\bibitem{Seto:2003kc} 
  N.~Seto and J.~'I.~Yokoyama,
  J.\ Phys.\ Soc.\ Jap.\  {\bf 72}, 3082 (2003)
  [gr-qc/0305096].
  
\bibitem{Boyle:2007zx}
L.~A.~Boyle and A.~Buonanno,
Phys.\ Rev.\ D {\bf 78}, 043531 (2008)
[arXiv:0708.2279 [astro-ph]].

  
\bibitem{Nakayama:2008ip} 
  K.~Nakayama, S.~Saito, Y.~Suwa and J.~'i.~Yokoyama,
  Phys.\ Rev.\ D {\bf 77}, 124001 (2008)
  [arXiv:0802.2452 [hep-ph]].
  
\bibitem{Nakayama:2008wy} 
  K.~Nakayama, S.~Saito, Y.~Suwa and J.~'i.~Yokoyama,
  JCAP {\bf 0806}, 020 (2008)
  [arXiv:0804.1827 [astro-ph]].

\bibitem{Kuroyanagi:2008ye} 
  S.~Kuroyanagi, T.~Chiba and N.~Sugiyama,
  Phys.\ Rev.\ D {\bf 79}, 103501 (2009)
  [arXiv:0804.3249 [astro-ph]].


\bibitem{Kuroyanagi:2009br} 
  S.~Kuroyanagi, C.~Gordon, J.~Silk and N.~Sugiyama,
  Phys.\ Rev.\ D {\bf 81}, 083524 (2010)
  [Erratum-ibid.\ D {\bf 82}, 069901 (2010)]
  [arXiv:0912.3683 [astro-ph.CO]].

\bibitem{Nakayama:2009ce} 
  K.~Nakayama and J.~'i.~Yokoyama,
  JCAP {\bf 1001}, 010 (2010)
  [arXiv:0910.0715 [astro-ph.CO]].

\bibitem{Kuroyanagi:2011fy} 
  S.~Kuroyanagi, K.~Nakayama and S.~Saito,
  Phys.\ Rev.\ D {\bf 84}, 123513 (2011)
  [arXiv:1110.4169 [astro-ph.CO]].

\bibitem{Kuroyanagi:2013ns} 
  S.~Kuroyanagi, C.~Ringeval and T.~Takahashi,
  Phys.\ Rev.\ D {\bf 87}, 083502 (2013)
  [arXiv:1301.1778 [astro-ph.CO]].



\bibitem{Turner:1993vb} 
  M.~S.~Turner, M.~J.~White and J.~E.~Lidsey,
  Phys.\ Rev.\ D {\bf 48}, 4613 (1993)
  [astro-ph/9306029].

\bibitem{Chongchitnan:2006pe} 
  S.~Chongchitnan and G.~Efstathiou,
  Phys.\ Rev.\ D {\bf 73}, 083511 (2006)
  [astro-ph/0602594].

\bibitem{Watanabe:2006qe} 
  Y.~Watanabe and E.~Komatsu,
  Phys.\ Rev.\ D {\bf 73}, 123515 (2006)
  [astro-ph/0604176].


\bibitem{Mendes:1998gr} 
  L.~E.~Mendes and A.~R.~Liddle,
  gr-qc/9811040.


\bibitem{Steigman:2012ve} 
  G.~Steigman,
  Adv.\ High Energy Phys.\  {\bf 2012}, 268321 (2012)
  [arXiv:1208.0032 [hep-ph]].

\bibitem{Tashiro:2003qp} 
  H.~Tashiro, T.~Chiba and M.~Sasaki,
  Class.\ Quant.\ Grav.\  {\bf 21}, 1761 (2004)
  [gr-qc/0307068].

\bibitem{AmaroSeoane:2012km} 
  P.~Amaro-Seoane, S.~Aoudia, S.~Babak, P.~Binetruy, E.~Berti, A.~Bohe, C.~Caprini and M.~Colpi {\it et al.},
  arXiv:1201.3621 [astro-ph.CO].

\bibitem{AmaroSeoane:2012je} 
  P.~Amaro-Seoane, S.~Aoudia, S.~Babak, P.~Binetruy, E.~Berti, A.~Bohe, C.~Caprini and M.~Colpi {\it et al.},
  Class.\ Quant.\ Grav.\  {\bf 29}, 124016 (2012)
  [arXiv:1202.0839 [gr-qc]].

\bibitem{bbo}
S. Phinney {\it et al.}, 
{\it The big bang observer: direct detection of gravitational waves from the birth of the Universe to the present}, 
NASA Mission Concept Study. 

\bibitem{Harry:2006fi}
 G.~M.~Harry, P.~Fritschel, D.~A.~Shaddock, W.~Folkner and E.~S.~Phinney,
 Class.\ Quant.\ Grav.\  {\bf 23}, 4887 (2006)
 [Erratum-ibid.\  {\bf 23}, 7361 (2006)].

\bibitem{Kawamura:2011zz} 
  S.~Kawamura, M.~Ando, N.~Seto, S.~Sato, T.~Nakamura, K.~Tsubono, N.~Kanda and T.~Tanaka {\it et al.},
  Class.\ Quant.\ Grav.\  {\bf 28}, 094011 (2011).


\end{thebibliography}
\end{document}